\documentclass[10pt,english]{revtex4}
\usepackage[T1]{fontenc}
\usepackage{CJKutf8}
\usepackage{color}
\usepackage{babel}
\usepackage{float}
\usepackage{units}
\usepackage{textcomp}
\usepackage{amsmath}
\usepackage{graphicx}
\usepackage{wasysym}
\usepackage[unicode=true,pdfusetitle,
 bookmarks=true,bookmarksnumbered=false,bookmarksopen=false,
 breaklinks=false,pdfborder={0 0 1},backref=false,colorlinks=false]
 {hyperref}

\makeatletter

\newcommand{\lyxmathsym}[1]{\ifmmode\begingroup\def\b@ld{bold}
  \text{\ifx\math@version\b@ld\bfseries\fi#1}\endgroup\else#1\fi}

\providecommand{\tabularnewline}{\\}

\@ifundefined{textcolor}{}
{%
 \definecolor{BLACK}{gray}{0}
 \definecolor{WHITE}{gray}{1}
 \definecolor{RED}{rgb}{1,0,0}
 \definecolor{GREEN}{rgb}{0,1,0}
 \definecolor{BLUE}{rgb}{0,0,1}
 \definecolor{CYAN}{cmyk}{1,0,0,0}
 \definecolor{MAGENTA}{cmyk}{0,1,0,0}
 \definecolor{YELLOW}{cmyk}{0,0,1,0}
}

\usepackage{babel}

\makeatother

\begin{document}
\title{Energetic, tunable, highly-elliptically polarized higher harmonics
generated by intense two-color counter rotating laser fields}
\author{E. Vassakis$^{1,2\dagger}$, S. Madas{\normalsize{}{}{}$^{3,4\dagger}$},
L. Spachis$^{2}$, T. Lamprou$^{1,2}$, I. Orfanos$^{1}$, S. Kahaly$^{3}$,
M. Upadhyay Kahaly$^{3*}$, D. Charalambidis$^{1,2,3}$ and E. Skantzakis$^{1}$}
\email{Corresponding author e-mail address: Mousumi.UpadhyayKahaly@eli-alps.hu, skanman@iesl.forth.gr}

\affiliation{$^{1}$Foundation for Research and Technology - Hellas, Institute
of Electronic Structure \& Laser, PO Box 1527, GR71110 Heraklion (Crete),
Greece}
\affiliation{$^{2}$Department of Physics, University of Crete, PO Box 2208, GR71003
Heraklion (Crete), Greece}
\affiliation{$^{3}$ELI-ALPS, ELI-HU Non-Profit Ltd., Wolfgang Sandner utca 3.,
Szeged, H-6728, Hungary}
\affiliation{$^{4}$Institute of Physics, University of Szeged, Dóm tér 9, H-6720
Szeged, Hungary}
\email{These authors contributed equally}

\begin{abstract}
In this work, we demonstrate experimentally the efficient generation
and tunability of energetic highly-elliptical high-harmonics in Ar
gas, driven by intense two-color counter rotating laser electric fields.
A bi-chromatic beam tailored by a MAZEL-TOV apparatus generates HHG,
where the output spectrum of the highly elliptical HHG radiation can
be tuned for a energy range of $\Delta$E$\HF$150 meV in the spectral
range of $\HF$20 eV with energy per pulse E$^{XUV}$$\HF$400 nJ
at the source. Furthermore we employ time-dependent density functional
simulations to probe in-depth the dependence of the harmonic ellipticity
and the strength of the isolated atto pulses on the driving field
parameters and demonstrate the robustness of the HHG with the bichromatic
field. We show how by properly tuning the central frequency of the
second harmonic, the central frequency of the extreme ultraviolet
(XUV) high harmonic radiation is continuously tuned. The demonstrated
energy values largely exceed the output energy from many other laser
driven attosecond sources reported so far and prove to be sufficient
for inducing nonlinear processes in the atomic system. We envisage
that such tunable energetic highly-elliptical HHG spectra can remove
the facility restrictions from requirements of few-cycle driving pulses
for isolated circular attosecond pulse generation. 
\end{abstract}
\maketitle

\section{Introduction\label{sec:Introduction}}

Ultrafast chiral processes can be studied using high harmonic generation
(HHG) \citep{Baykusheva2018,Ayuso2019,Cireasa2015,Heinrich2021,Neufeld2018,Neufeld2019}.
HHG is an extreme nonlinear process, which depends on many parameters,
such as the non-linear medium, the phase matching conditions, the
peak intensity, the duration and the spectral phase of the driving
pulse. By controlling these parameters it is possible to manipulate
the spectral characteristics of the emitted extreme ultraviolet (XUV)
radiation \citep{Altucci1996,Chang1998,Lee2001,Winterfeldt2008}.
It has been demonstrated experimentally and theoretically that exploiting
different mechanisms the harmonic HHG spectra can be influenced. Controlling
parameters include the temporal chirp of the driving laser, the ionization-induced
blue shift of the driver pulse in the generating medium \citep{Froud2006,Altucci1996,Winterfeldt2008}
and tuning the central wavelength of the driver using an optical parametric
amplifier \citep{Shan2002}. HHG from linear polarized drivers produces
harmonics with linear polarization and therefore severe limitations
are imposed to the potential applications of an XUV HHG source \citet{Chordiya2023}.
In this context, the development of a tunable highly-elliptical XUV
source adds an advanced tool. Such sources have recently emerged as
a central topic of ultrafast science promising invaluable insights
into chiroptical phenomena taking place on ultrashort time scales.
Ultrashort circularly polarized pulses in the XUV domain have been
generated at large-scale facilities, such as free-electron lasers\citep{McNeil2010,Higley2016,Lutman2016}
and femto-sliced synchrotrons \citep{Khan2006,Yamamoto2011,ifmmodecheckCelsevCfiutiifmmodeacutecelsecfi2011}.
In an effort to make such sources more broadly available, table-top
sources based on high-harmonic generation (HHG) have also been developed
\citep{Kfir2015,Comby2020,Ferre2015,Hickstein2015,Vassakis2021,Comby2022}.

The energy content of laser driven highly-elliptical or circularly
polarized XUV radiation was limited to the pJ range per pulse \citep{Kfir2015,Comby2020,Ferre2015,Hickstein2015}
until recently when energy content in the nJ range were demonstrated
in ref.\citep{Vassakis2021}. It is known that the recollision mechanism,
which describes the HHG, entails that a stronger chiral response arises
at the cost of a greatly suppressed high harmonic emission signal
\citep{Cireasa2015}. Thus, the generation of energetic highly elliptical
high harmonics is also a decisive step towards chiral-matter investigations.

In the present work, tunable energetic highly-elliptical HHG in the
XUV regime is theoretically studied and experimentally demonstrated,
exploiting two-color counter rotating electric fields under loose
focusing geometry. The output spectrum of the highly elliptical HHG
radiation can be tuned for an energy range of $\Delta$E$\HF$150
meV in the spectral range of $\HF$20 eV with energy per pulse E$^{XUV}$$\HF$400
nJ at the source.This is to our knowleeedge, the highest reported
energy content per laser pulse in the laser-driven highly elliptical/circularly
polarized XUV radiation.

\begin{figure*}[t]
\begin{centering}
\includegraphics[width=1\textwidth]{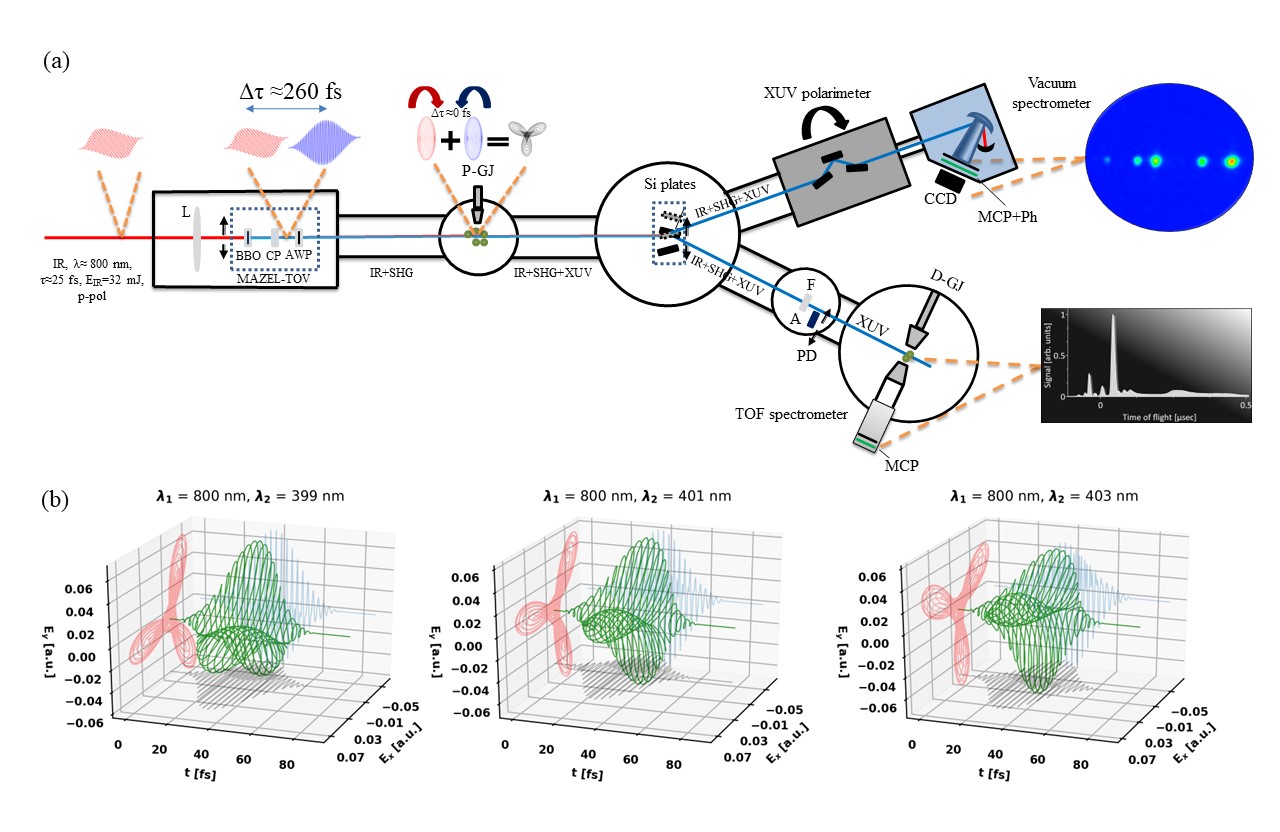} 
\par\end{centering}
\caption{(a) Experimental apparatus for the generation of highly elliptical
energetic tunable XUV radiation . A compact (15 cm long) MAZEL-TOV-like
device is installed after a 3-m focal length lens (L). The device
includes a BBO crystal, a calcite plate (CP) (both mounted independently
on rotatable stages with high precision) and a rotatable super achromatic
quarter waveplate(AWP). The two-color bi-circular field beam is focused
into a pulsed gas jet filled with Ar (P-GJ). The generated XUV radiation
is reflected towards the detection area by Si plates which consists
of two branches. The first branch hosts a calibrated XUV photodiode
(PD), a pulsed gas jet filled with Ar (D-GJ) and finally a $\mu$-metal
shielded TOF spectrometer. The second branch hosts a rotating in-vacuum
polarizer and consequently, the XUV radiation is diffracted by a spherical
holohgraphic grating and detected by a microchannel-plate (MCP) detector
coupled to a phosphor screen. (b) Calculated Bi-circular counter rotating
electric field formed at the output of MAZEL-TOF-like device, based
on our experimental parameters, with fundamental field having a central
wavelength "$\lambda_{1}$" of 800 nm and three different second
harmonic field wavelengths "$\lambda_{2}$", i.e., 399 (a), 401
(b), and 403 nm \label{fig:Experimental-apparatus-for}}
\end{figure*}

\section{Adopted methodologies: experiments and simulations\label{sec:Experiment}}

For the experimental implementation, a compact MAch-ZEhnder-Less for
Threefold Optical Virginia spiderwort-like (MAZEL-TOV-like \citep{Kfir2016})
scheme is used to generate elliptically polarized HHG spectra. The
technique is detailed in subsection 2.1, and schematically described
in Fig. 1a. to In order to understand the spectral and temporal structure
of the generated high-order harmonics our experimental characterizations
and results are supported by theoretical simulations based on semi-classical
approach, as elaborated in subsection 2.2 and time-dependent density
functional theory in subsection 2.3, sub-cycle dynamics and crucial
trends, identifying the factors in the estimated ellipticity of different
harmonics.

\subsection{Experimental Setup\label{subsec:Experimental-Setup}}


The experiment was performed by exploiting the MW beamline of the
Attosecond Science and Technology Laboratory (AST) at FORTH-IESL.
The experiment utilizes a 10 Hz repetition rate Ti:Sapphire laser
system delivering pulses with up to 400 mJ/pulse energy, $\tau_{L}$=25
fs duration and a carrier wavelength of 800 nm. The experimental set-up
consists of three areas: the focusing and MAZEL-TOV-like \citep{Kfir2016}
device chamber, the harmonic generation chamber and the detection
chambers (Fig.\ref{fig:Experimental-apparatus-for}). A laser beam
of 3 cm outer diameter and energy of 32 mJ/pulse is passing through
a 3 m focal-length lens with the MAZEL-TOV-like device positioned
1.25 meters downstream. The apparatus consists of a beta-phase barium
borate crystal (BBO), a calcite plate and a super achromatic quarter
waveplate. A fraction of the energy of the linear p-polarized fundamental
pulse, is converted into a perpendicularly polarized (s-polarized)
second harmonic field in a BBO (0.2 mm, cutting angle $29.20^{o}$
for type I phase matching). The conversion efficiency of the BBO crystal
was maximized and it was found \ensuremath{\approx} 30\% at 403 nm.
The run-out introduced by the BBO crystal for the SHG of 800 nm was
determined to be 38.6 fs. It is noted that by placing the BBO after
the focusing lens ensures that the wavefronts of the converging fundamental
laser beam are reproduced into that of the second harmonic field.
Therefore, the foci (placed close to a pulsed gas jet filled with
Ar) of the $\omega$ and 2$\omega$ fields coincide along the propagation
axis. Additionally, the beam passes through a calcite plate at almost
normal incidence (AR coated, group velocity delay (GVD) compensation
range 310-450 fs), which pre-compensates group delays introduced by
the BBO crystal and the super achromatic quarter waveplate. The super
achromatic waveplate converts the two-color linearly polarized pump
into a bi-circular field, consisting of the fundamental field and
its second harmonic, accumulating at the same time a group delay difference
of $\lyxmathsym{\AC}$253 fs between the $\lyxmathsym{\AC}$400 nm
and 800 nm central wavelengths. Assuming Gaussian optics, the intensity
at the focus for the two components of the bi-circular polarized field
is estimated to be $I_{\omega}$\ensuremath{\approx}$I_{2\omega}$\ensuremath{\approx}
1x10$^{14}$ W/cm$^{2}$. After the jet, the produced XUV co-propagates
with the bi-circular driving fields towards a pair of Si plates, which
are placed at 75$^{o}$ reducing the p-polarization component of the
fundamental and the second harmonic radiation while reflecting the
harmonics \citep{Takahashi2004} towards the detection area. The detection
area conctinst of two branches. In the first branch which is placed
directly after the first Si plate, a pair of 5 mm diameter apertures
were placed in order to block the outer part of the $\omega$ and
2$\omega$ beams, while letting essentially the entire XUV through.
A 150 nm thick Sn filter is attached to the second aperture, not only
for the spectral selection of the XUV radiation, but also to eliminate
the residual bicircular field. A calibrated XUV photodiode (XUV PD)
can be introduced into the beam path in order to measure the XUV pulse
energy. The transmitted beam enters the detection chamber, where the
spectral characterization of the XUV radiation takes place. The characterization
is achieved by recording the products of the interaction between the
XUV beam and the Ar atoms. The electrons produced by the interaction
of Ar atoms with the unfocused XUV radiation were detected by a $\mu$-metal
shielded time-of-flight (TOF) spectrometer. The spectral intensity
distribution of the XUV radiation is obtained by measuring the single-photon
ionization photo-electron (PE) spectra induced by the XUV radiation
with photon energy higher than the $I_{p}$ of Ar ($I_{pAr}$=15.76
eV). In the second branch and after the Si plate, a rotating in-vacuum
polarizer is installed and consequently, the XUV radiation is diffracted
by a spherical holographic grating and detected by a microchannel-plate
detector (MCP) coupled with a phosphor screen.

\subsection{Simulations}

\subsubsection{Semi-classical approach}

In order to have an intuitive picture of the HHG by two color bi-circular
polarized fields, we performed calculations based on the semicalssical
approach where the theoretical framework and detailed analysis can
be found elsewhere \citep{Miloifmmodecheckselsevsfieviifmmodeacutecelsecfi2000}.
In our supplementary material \ref{subsec:The-Semiclassical-analysis},
we present an abbreviated analysis which is based on Strong Field
Approximation (SFA)\citep{Lewenstein1994} adapted in the case of
these electric fields.

Within this model, the radiation at orders 3n$\pm1$ (with n=1,2,3,...),
emitted from a single atom exposed to a intense bi-chromatic driving
electric field \textbf{\textcolor{black}{E(t)}} with the associated
vector potential A(t) = \textminus{} $\varint\mathbf{E}(t)dt$ can
be fully characterized by the Fourier transform of the time-dependent
dipole moment. Details of this analysis can be found elsewhere \citep{Nayak2019}.

\subsubsection{Time dependent density functional formalism}

The time-dependent electron dynamics in Argon under the influence
of bi-chromatic counter rotating (BCCR) laser fields are investigated
based on \textit{ab-initio} calculation within time dependent density
functional theory (TDDFT) approach \citep{Runge1984} in the real-time
and real-space grids as implemented in the octopus computational package
\citep{Andrade2012,Andrade2015}. Thus, a more detailed and a more
complete picture of these dynamics are presented.

The driving laser fields, as described in terms of an electric field,
are polarized in the x-y plane and dipole approximation is considered.
In addition, the contributions from magnetic component of the electromagnetic
field and any other relativistic terms, such as spin-orbit coupling
are neglected. The combined electric field is a superposition of two
counter rotating laser fields, and is given as in \ref{rosette}.

\begin{equation}
\begin{split}E(t)=\sum_{i=1,2}E_{i} & \cos{\bigg(\frac{(t-\tau_{\omega1})}{\tau_{\omega i}}\bigg)^{2}}\Bigg[\cos{\bigg(\omega_{i}(t-\tau_{\omega1})\bigg)}\pmb{e_{x}}+\\
 & a_{i}\sin{\bigg(\omega_{i}(t-\tau_{\omega1})\bigg)}\pmb{e_{y}}\Bigg]
\end{split}
\label{rosette}
\end{equation}

where \textbf{$e_{x}$} and \textbf{$e_{y}$} are the two mutually
perpendicular unit vectors. We considered two counter rotating fields
of $a_{1}$ = $-a_{2}~=~1$ with $E_{1}~=~E_{2}~=E_{0}$. The peak
laser intensity is expressed in terms of field strength $E_{0}:I=E_{0}^{2}I_{a}$,
where $I_{a}=3.51\times10^{16}~W/cm^{2}$ is the atomic intensity
unit. The peak intensity is $I=1\times10^{14}~W/cm^{2}$. $\cos{\bigg(\frac{(t-\tau_{\omega1})}{\tau_{\omega i}}\bigg)^{2}}$
denotes the envelope of the pulse, where $\tau_{\omega1}$ is the
total duration of the fundamental pulse, that may be defined in terms
of full width at half maximum of intensity (FWHM), $\tau_{p}:\tau_{\omega1}=\tau_{p}/(2\arccos{(2^{-1/4})})$,
where $\tau_{p}=25$ fs. $\tau_{\omega2}$ is the pulse duration of
the second harmonic field, which is defined as $\tau_{\omega2}=\tau_{\omega1}/\sqrt{2}$.
A central wavelength $\lambda_{1}=800$ nm is used for the fundamental
field and four different wavelengths are used for the second harmonic
field, which are $\lambda_{2}=399,~\text{or}~400,~\text{or}~401,~\text{or}~403$
nm. 

The initial states are obtained from self-consistent solutions of
wave functions at the density functional theory (DFT) level. Later,
those states are propagated by using the approximated enforced time-resolved
symmetry (AETRS) method, with $\Delta t~=~0.3$ a.u. time-steps. Rest
of the TDDFT input parameters are presented in the supplementary material
(\ref{TDDFT_suppl}). The harmonic spectral properties are calculated
from the resulting dipole acceleration signal, which has components
that are both parallel and perpendicular to the laser polarization.
The harmonic spectrum $HHG(\omega)$ is obtained from the Fourier
transformation of the time-dependent dipole acceleration \textbf{a(t)}
and can be written as \citep{Chu2001,Burnett1992},

\begin{equation}
HHG(\omega)=\bigg|\mathcal{FT}\bigg(\int_{-\infty}^{\infty}{a(t)~dt}\bigg)\bigg|^{2},
\end{equation}

where, $\mathcal{FT}$ is the Fourier transform.

Three of the different bichromatic circularly polarized field variations
having coplanar counter-rotating components, that are used in both
experiment and simulations in our work, are shown in Fig. \ref{fig:Experimental-apparatus-for}(b).

\section{Results and analysis}

\subsection{Highly-elliptical XUV radiation spectral characteristics}

\label{subsec:Highly-elliptical-XUV-radiation}

\begin{figure}[h]
\begin{centering}
\includegraphics[width=0.5\textwidth]{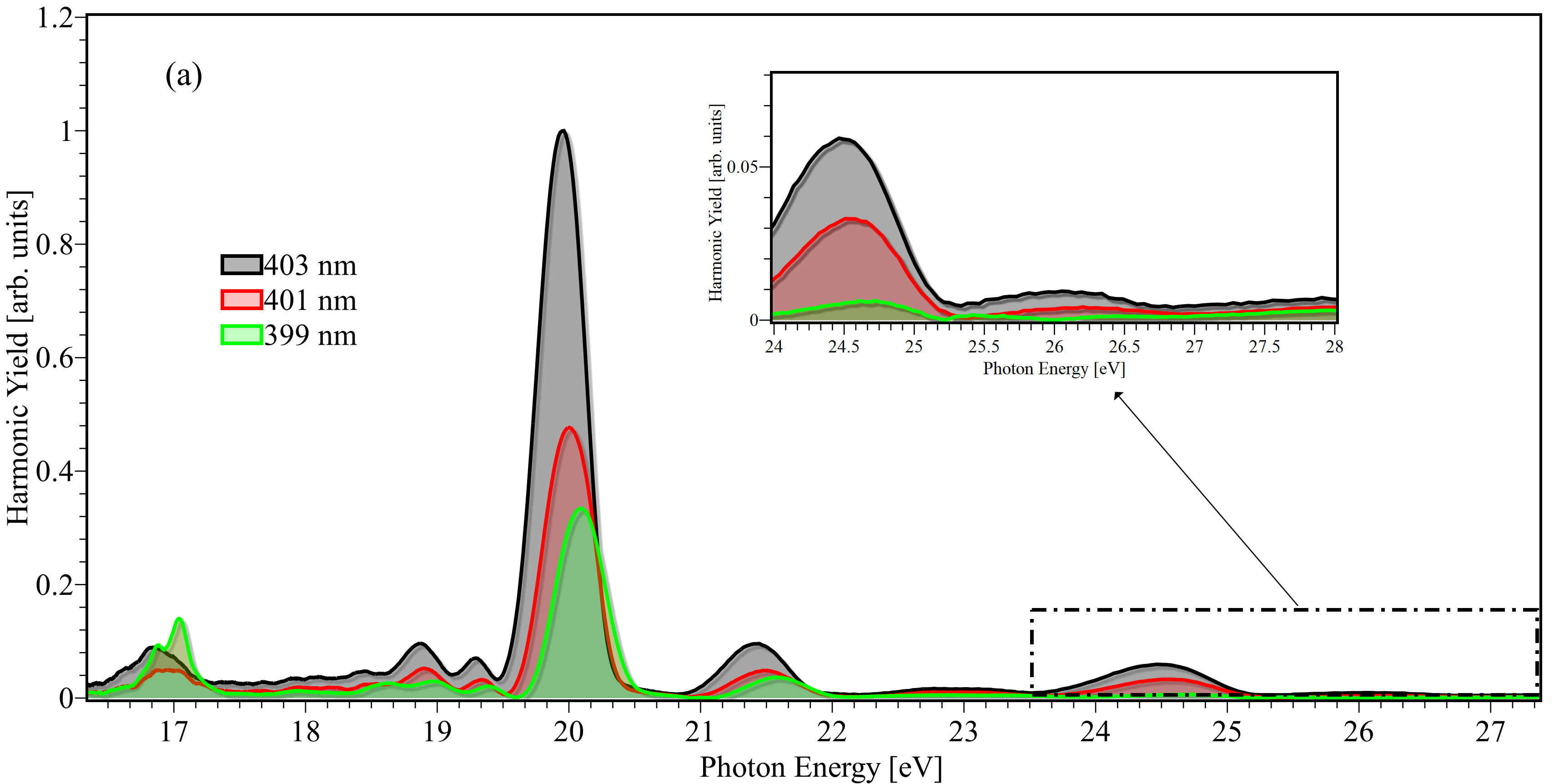} 
\par\end{centering}
\begin{centering}
\includegraphics[width=0.25\textwidth]{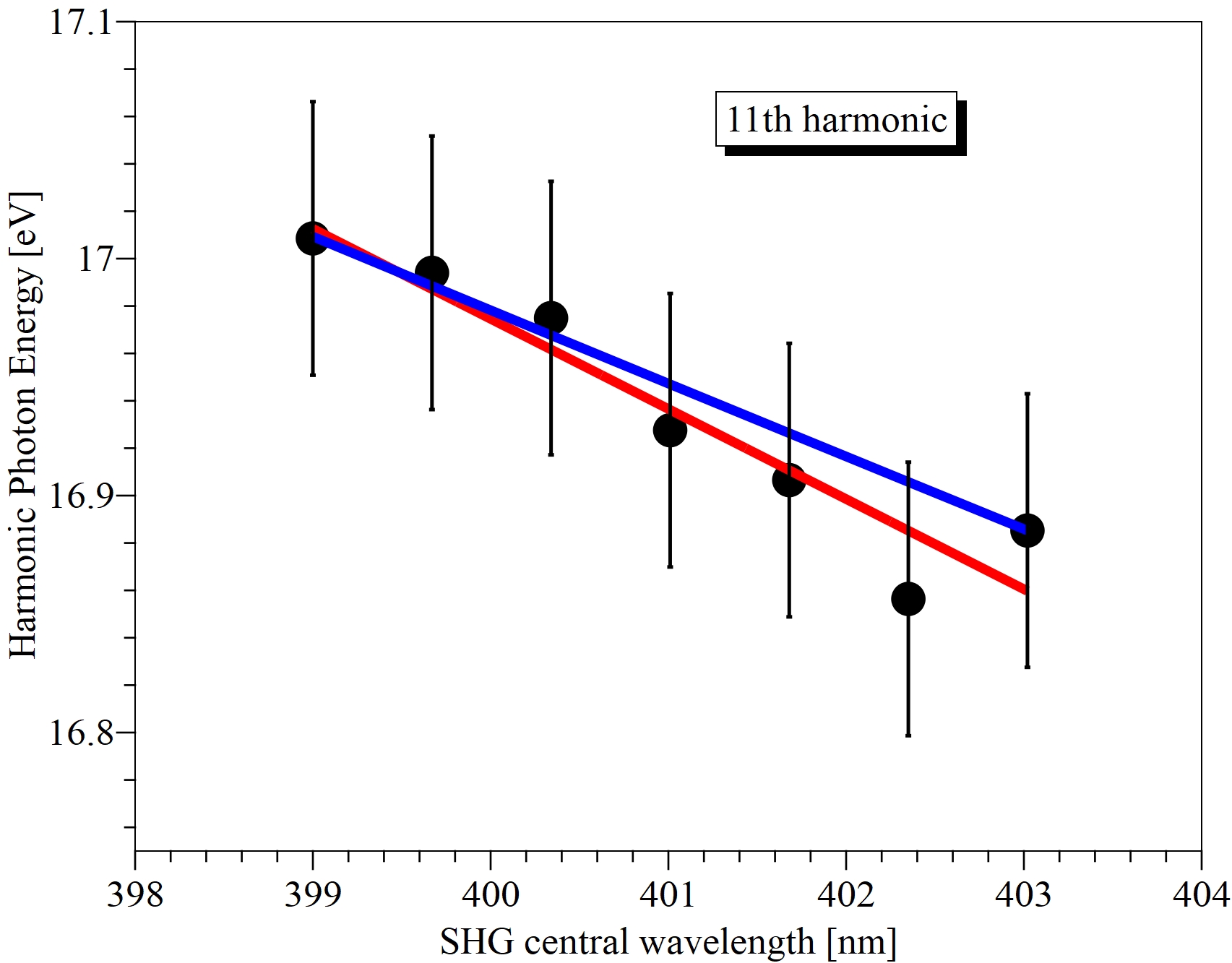}\includegraphics[width=0.25\textwidth]{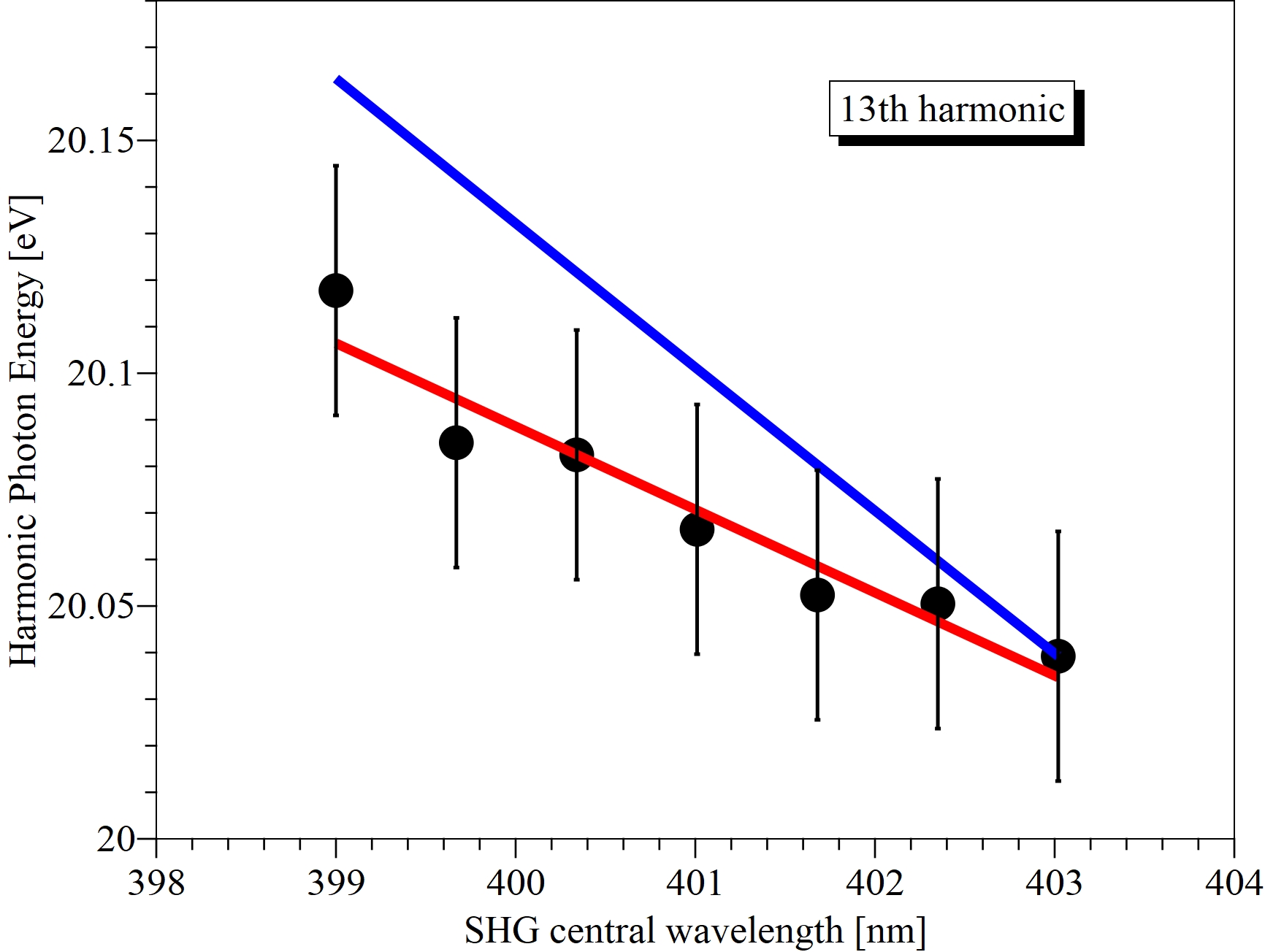} 
\par\end{centering}
\begin{centering}
\includegraphics[width=0.25\textwidth]{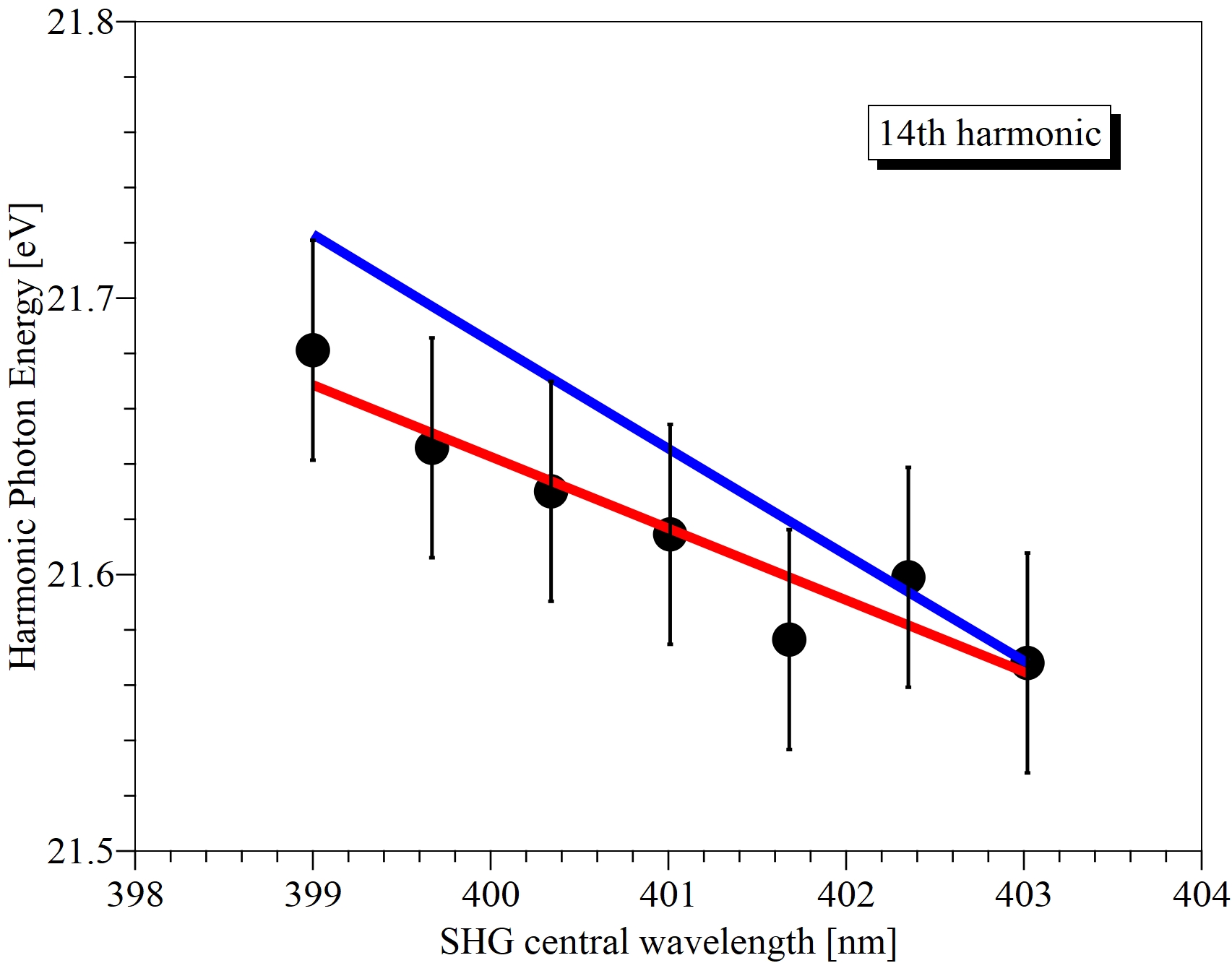}\includegraphics[width=0.25\textwidth]{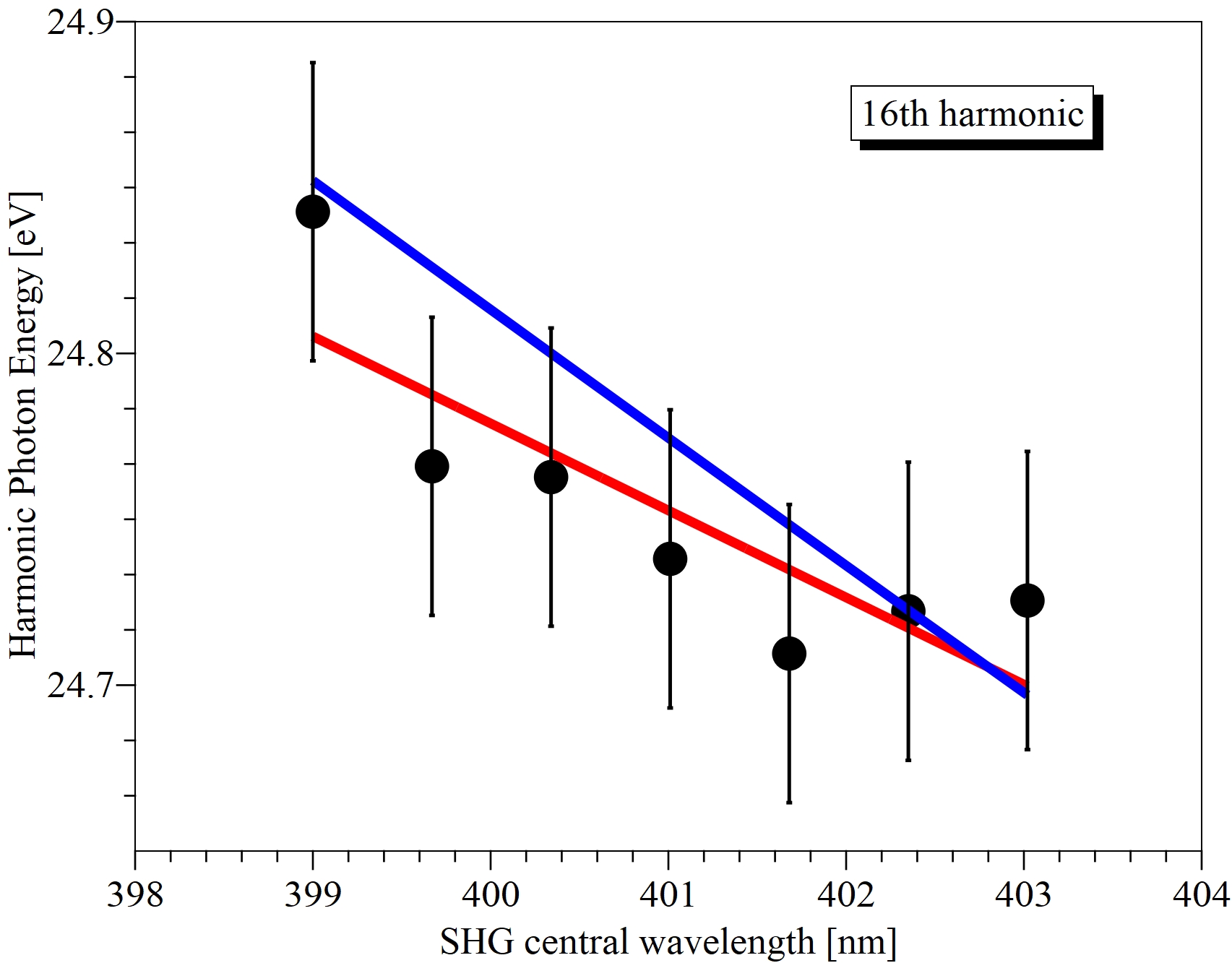} 
\par\end{centering}
\begin{centering}
\includegraphics[width=0.5\textwidth]{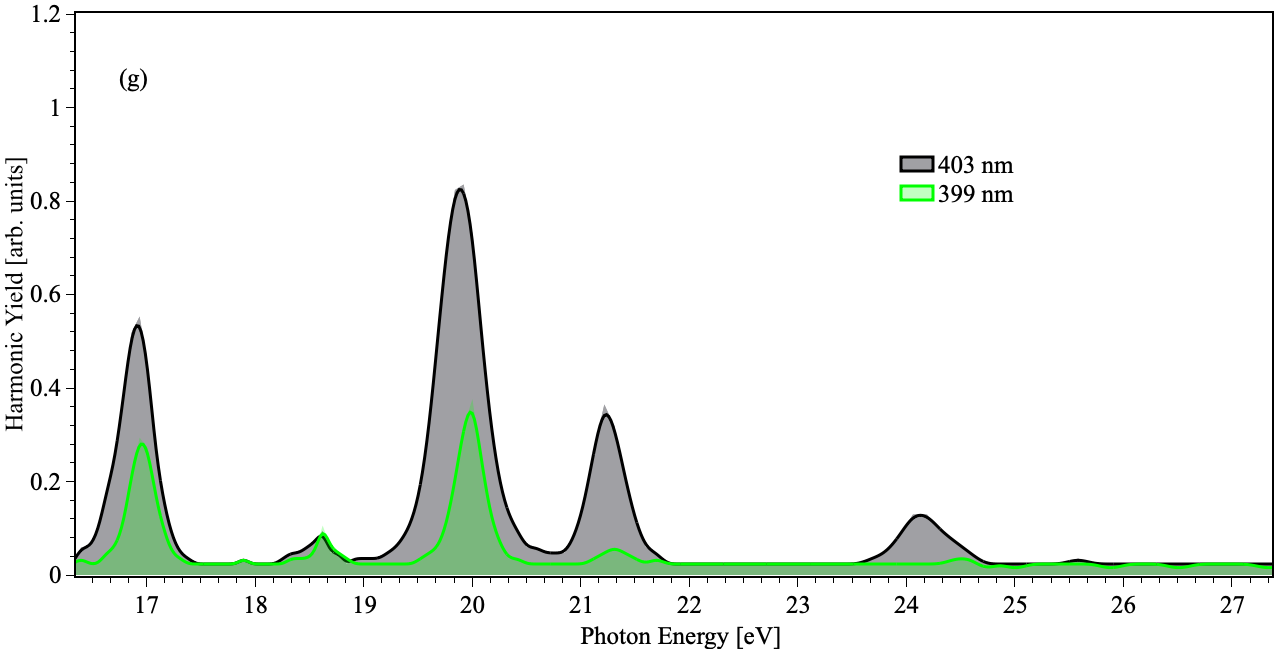} 
\par\end{centering}
\caption{(a) Characteristic highly-elliptical HHG spectra for three different
values of central SHG wavelength. The maximum energy shift observed
is in the order of $\Delta$E$\protect\HF$150 meV. (b),(c),(d),(e)
Complementary analysis of the detectable harmonics (11$^{th}$,13$^{th}$,14$^{th}$
and 16$^{th}$). An linear dependence of central HHG energy on the
central wavelength of the SHG is observed. The blue line depicts the
same dependence calculated from the energy conservation of the annihilated
driver photons and the emitted XUV photon\label{fig:HHG-Yield-vs}.
Figure \ref{fig:HHG-Yield-vs}(g) presents highly-elliptical HHG spectra
obtained by the vacuum spectrometer.}
\end{figure}

For the spectral characterization of the highly-elliptical XUV radiation
different PE spectra were recorded as a function of the angle \ensuremath{\Delta}$\theta$
of the BBO crystal installed in the MAZEL-TOV-like device. Characteristic
highly-elliptical HHG spectra are presented in Fig.\ref{fig:HHG-Yield-vs}(a)
for three different positions of \ensuremath{\Delta}$\theta$. An
energy shift is clearly observed towards higher photon energies when
the the angle (\ensuremath{\Delta}$\theta$) between the propagation
axis of the IR driving field and the BBO crystal is increased. Additionally,
in Fig.\ref{fig:HHG-Yield-vs}(b),(c),(d),(e) complementary measurements
of the detectable harmonics by the TOF spectrometer are presented.
Finally, Fig. \ref{fig:HHG-Yield-vs}(g) presents highly-elliptical
HHG spectra obtained by the vacuum spectrometer for reason of completeness.
It is revealed that central HHG energy indicating almost linear dependence
on the central wavelength of the SHG. The maximum energy shift observed
is in the order of $\Delta$E$\HF$150meV. The observed blue shift
can be attributed to the energy conservation of the HHG process by
two color BCCR electric fields. From energy conservation, the harmonic
frequencies generated by these fields is given as
\begin{equation}
\varOmega_{(n_{1},n_{2})}=n_{1}\cdot\omega+n_{2}\cdot\beta\omega
\end{equation}
where, $n_{1}$ and $n_{2}$ are integer numbers that are associated
with the number of photons involved in the process of HHG at angular
frequencies $\omega$ and $\beta\omega$, respectively. On the other
hand, parity and spin angular momentum conservation requires that
$\varDelta n=n_{1}-n_{2}$=$\pm1$. Therefore, in the case of 13$^{th}$
harmonic and $1.985\leq\beta\leq2.005$ (which reflects our experimental
conditions), the unique pair ($n_{1},n_{2}$) associated with this
harmonic is equal to (5,4). It becomes evident that a linear dependence
of the generated harmonic photon energy is expected as a function
of the parameter $\beta$ (as confirmed in Table \ref{tab:The-slope-=00003D00003D000394=00003D00003D000395/=00003D00003D000394=00003D00003D0003BB}).

In Fig.\ref{fig:HHG-Yield-vs}(b),(c),(d),(e), is depicted with the
blue line this dependence calculated from the energy conservation
of the annihilated driver photons and the emitted XUV photons (Table
\ref{tab:The-slope-=00003D00003D000394=00003D00003D000395/=00003D00003D000394=00003D00003D0003BB})\citep{Fleischer2014}

\begin{table}[h]
\begin{centering}
\begin{tabular}{|c|c|c|}
\hline 
\textbf{harmonic}  & \textbf{(n$_{1}$,n$_{2}$)}  & \textbf{$\Delta E/\Delta\lambda$ {[}eV/nm{]}}\tabularnewline
\hline 
\hline 
11$^{th}$  & (3,4)  & -0,03086\tabularnewline
\hline 
13$^{th}$  & (5,4)  & -0,03086\tabularnewline
\hline 
14$^{th}$  & (4,5)  & -0,03858\tabularnewline
\hline 
16$^{th}$  & (6,5)  & -0,03858\tabularnewline
\hline 
17$^{th}$  & (5,6)  & -0,04656\tabularnewline
\hline 
\end{tabular}
\par\end{centering}
\caption{The slope $\Delta E/\Delta\lambda$ {[}eV/nm{]} respect to each harmonic
order. For each harmonic order, the corresponding channel is also
displayed, where n$_{1}$ is the number of photons of the fundamental
field and n$_{2}$ is the number of photons of the second harmonic
field.\label{tab:The-slope-=00003D00003D000394=00003D00003D000395/=00003D00003D000394=00003D00003D0003BB}}
\end{table}

From another perspective, this characteristic blue shift can be interpreted
in the framework of Strong Field Approximation (SFA)\citep{Lewenstein1994,Miloifmmodecheckselsevsfieviifmmodeacutecelsecfi2000}.
The active electron acquires different complex phase in the continuum
for each central SHG wavelength. This leads to constructive interference
at different spectral positions inside the highly-elliptical HHG spectrum
resulting to the observed blue shift in the energy domain.

\subsection{Polarimetric investigation \label{subsec:Polarimetric-characterization-of}}

\begin{figure}[h]
\centering{}\includegraphics[width=1\textwidth]{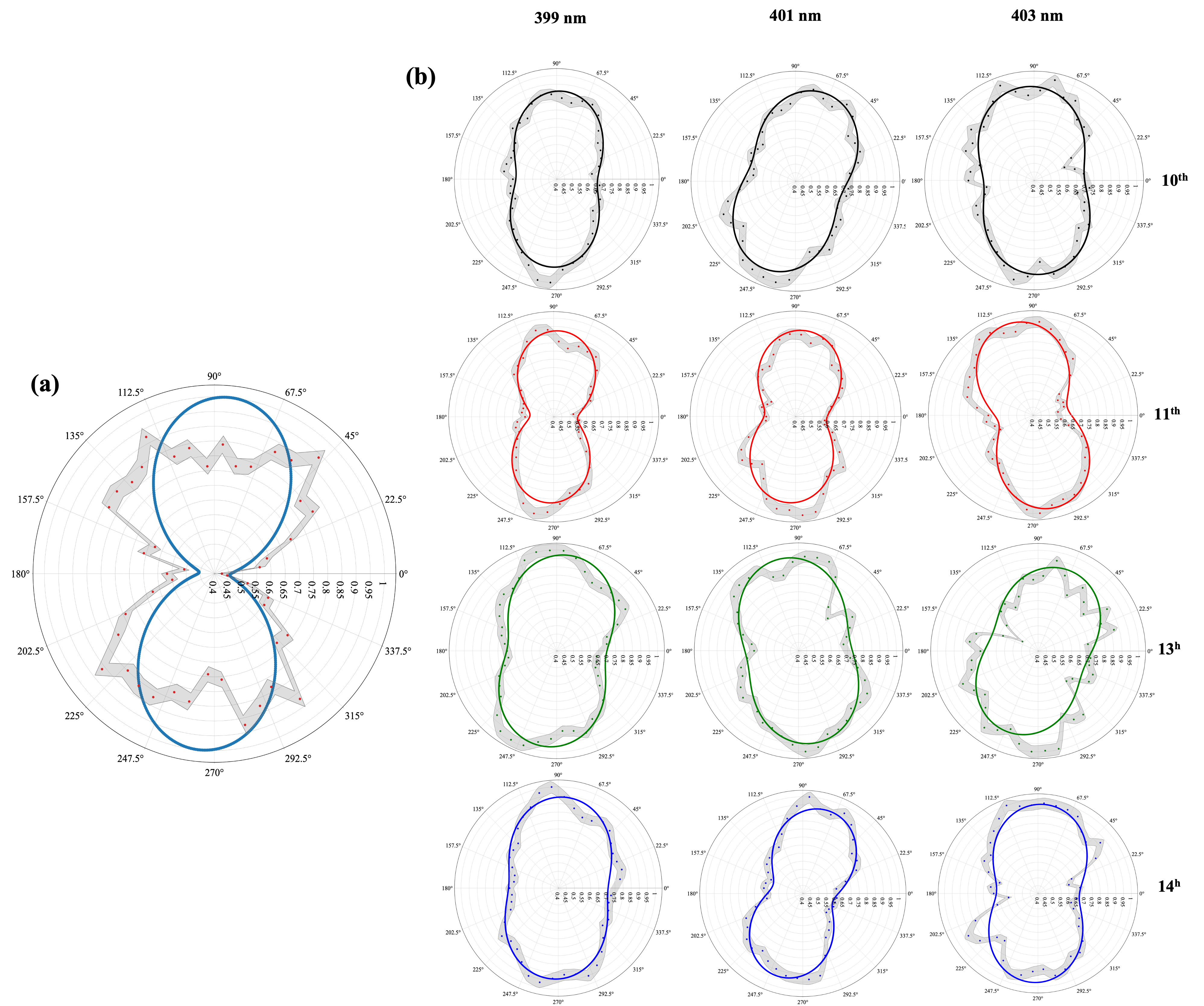}\caption{(a) Characteristic polarimetric measurment for the 11$^{th}$ harmonic
of the linear polarized driver. (b) Polarization scan for the 10$^{th}$,
11$^{th}$, 13$^{th}$, 14$^{th}$ harmonics using for three different
values of the BBO turning angle $\Delta$$\vartheta$ with respect
to the propagation axis of the 25-fs IR laser pulse of the driver
and thus different phase matching SHG central wavlelengths.\label{fig:Polarimetry}}
\end{figure}

The emitted harmonic radiation after the reflection of a Si plate
is sent through a rotating in-vacuum polarizer, which consists of
two UV protected silver(Ag)-coated and one UV protected aluminum-coated(Al)
mirrors mounted on a common plate. The angle of incidence on each
mirror is $75^{o}(Ag)$-$60^{o}(Al)$-$75^{o}(Ag)$ . Finally, the
harmonic radiation is diffracted using a spherical holographic grating
(platinum coated, 2400 G/mm) and detected by a multichannel-plate
detector (MCP) coupled to a phosphor screen. The contrast of the rotating
in-vacuum polarizer, was extracted by polarimetric measurements of
the linear p-polarized XUV radiation (MAZEL-TOV-like device out of
the beam line) which is presented in Fig. \ref{fig:Polarimetry}(a)
and it was found to be R = 2.2 $\pm$ 0.1 by fitting the imperfect
polarizer equation \ref{eq:26} (For the mathematical extraction of
the imperfect polarizer equation see Section \ref{subsec:Polarization---Reflective})
. Polarimetric measurements were performed for the 10$^{th}$, 11$^{th}$,
13$^{th}$, 14$^{th}$ harmonics as a function of the angle \ensuremath{\Delta}$\theta$.
Therefore, polarization scans were acquired during which the polarizer
was rotated between 0° and 360° in step of $\sim$8° and averaging
of 40 pulses per step.

\begin{figure}[h]
\begin{centering}
\includegraphics[width=0.5\textwidth]{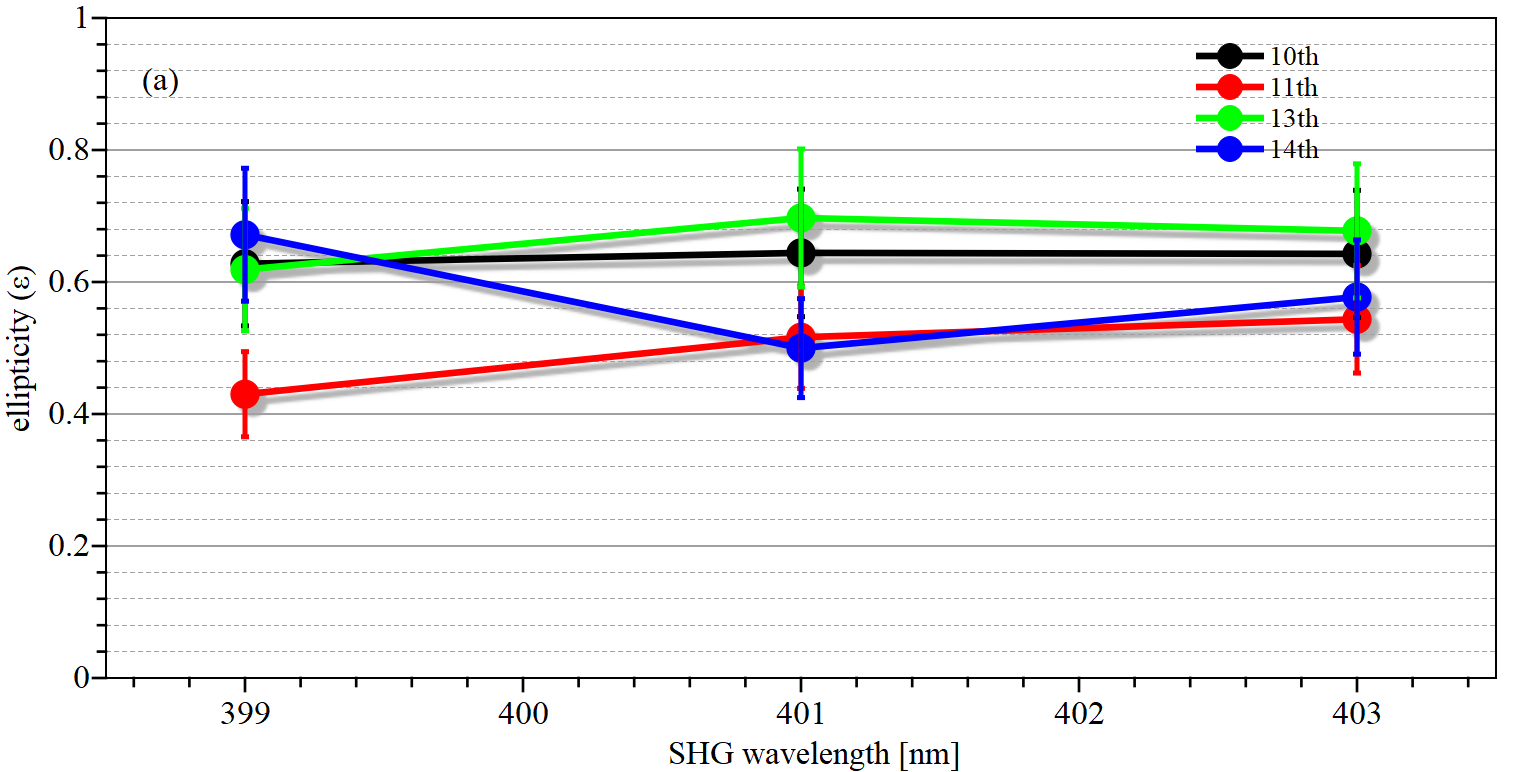}\includegraphics[width=0.5\textwidth]{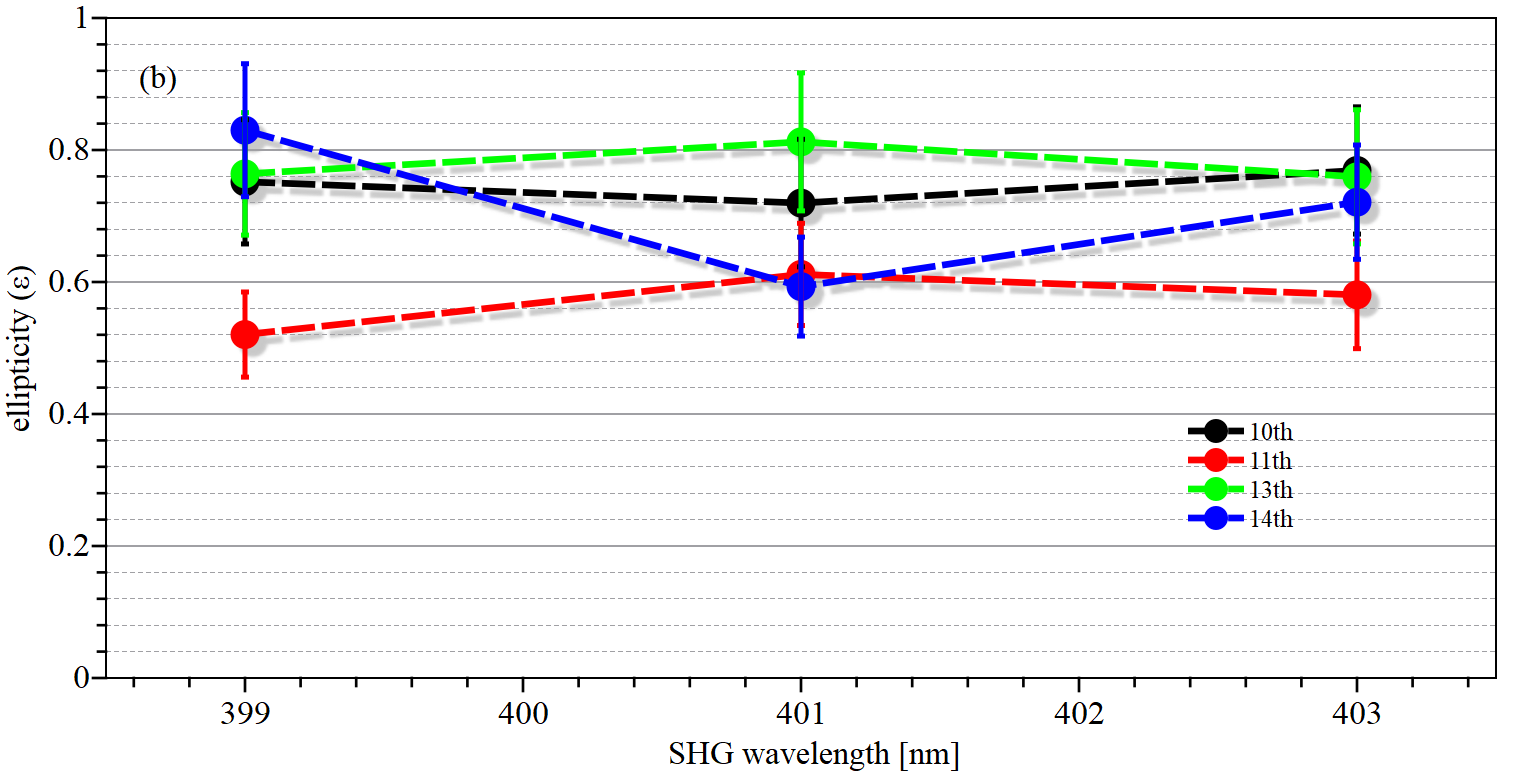} 
\par\end{centering}
\centering{}\caption{(a)The dependence of ellipticity on the central wavelegth of the SHG
of the fundamental frequency at the \emph{detection} area and (b)the
reconsructed ellipticity values at the \emph{source} .\label{fig:ellipticities}}
\end{figure}

From these scans the ellipticity of the high harmonics as a function
of the BBO angle was derived. The measurement of the state of polarization
of the HHG spectra confirmed highly-elliptical polarization reaching
ellipticities up to $\sim70$\% at $\sim$22 eV as it is presented
in Fig. \ref{fig:ellipticities}(a). No dependence on the central
wavelength of the SHGwas found. Briefly, the polarization state of
these highly elliptical harmonics was extracted by fitting the imperfect
polarizer equation (\ref{eq:26} of the Supplementary Material) on
the raw data of the polarization scans keeping as only known parameter
the contrast of the polarizer R = 2.2 $\pm$ 0.1. In the present configuration
used for the characterization of the harmonic ellipticity includes
the Si plate which reflects highly elliptical harmonic emission to
the XUV polarimeter. This optical component provides marginally different
reflectivity for the s and p polarization component of the XUV radiation
affecting also the measured ellipticity of the HHG radiation. By calculating
the reflectivity of the Si plate for the two polarization orientations
by using the corresponding Fresnel equations using the refractive
indexes taken from Palik et al.\citep{Palik1997} and for each harmonic
component the values of ellipticity at the \emph{source} can be estimated.
Therefore, ellipticities up to $\sim85$\% at $\sim$22 eV were found
as it is shown in Fig. \ref{fig:ellipticities}(b). These ellipticity
values in conjuction with the high energy content of the XUV pulses,
reveal a source adequate for in applications like control and imaging
of ultrafast magnetism in magnetic materials \citep{Kfir2017,Jilili2023,Siegrist2019},
in ultrafast chiral matter investigations \citep{Ferre2015} and also
in the investigation of circular dichroism in atomic systems \citep{Hofbrucker2018,Lambropoulos1972,Mayer2022}
where intense highly-elliptical/cicrcularly polarized XUV radiation
is necessary for inducing nonlinear processes\citep{Lambropoulos1972}.

\subsection{Energy Content estimation\label{subsec:Energy-Content-estimation}}

\begin{figure}[h]
\centering{}\includegraphics[width=1\textwidth]{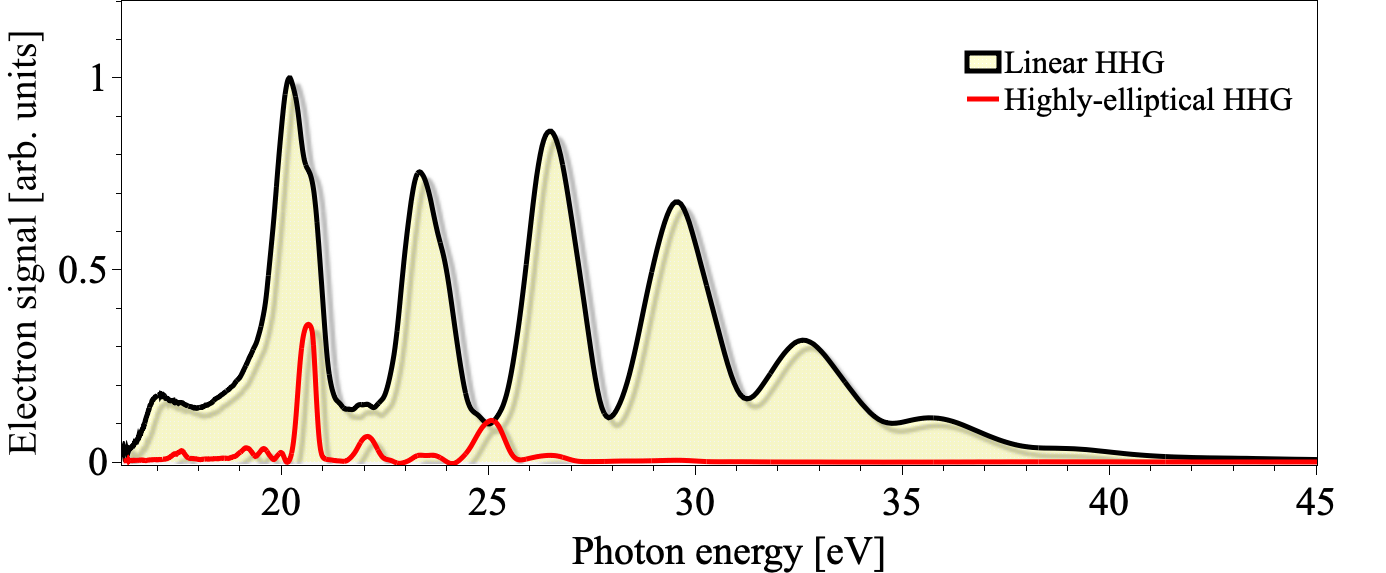}\caption{Typical PE spectra of Ar induced by the linear p-polarized and highly-elliptical
XUV radiation generated by Ar at the same detection conditions, after
the optimization of harmonic emission is realized in both cases.\label{fig:Typical-PE-spectra}}
\end{figure}

The energy content estimation of the XUV radiation was estimated by
applying the experimental procedure similar to the one reported in
Vassakis et al. \citep{Vassakis2021} and it was found to be E$^{XUV}$$\HF$400
nJ per XUV pulse at the source. Briefly, the energy of the highly
elliptical XUV radiation emitted per pulse is estimated at a first
step by measuring the linearly polarized XUV pulse energy by means
of an XUV photodiode and comparing the HHG spectra depicted in the
measured photoelectron spectrum of Ar atoms upon interaction with
highly elliptically and linearly polarized XUV light (see Fig. \ref{fig:Typical-PE-spectra})
at the same detection conditions. Additionally, the single-photon
photoelectron angular distributions were taken into account for each
highly elliptical harmonic in the case of Ar\citep{Schoenhense1980}
and in order to check how they affect the detection in the case of
our experimental configuration. From all the above, we can deduce
the energy per pulse of the highly-elliptical XUV emission. This major
improvement with respect to the previous reported value ($\HF$100
$nJ$ at the source)\citep{Vassakis2021} was due to the improvement
of the focusing conditions and the increased input laser energy (32
mJ) of the TW 10 Hz laser system. It should be stressed that no depletion
of the atomic target (Ar) was observed during these experimental investigations
and therefore higher energy content of highly elliptical XUV radiation
is expected by increasing the input energy, therefore the limit for
the driving laser energy was set by the possible damage threshold
of the optical components. The estimated number of photons per harmonic
per pulse at the source is shown in Table \ref{Table Number of Photons}

\begin{table}[H]
\centering{}%
\begin{tabular}{|c|c|}
\hline 
Harmonic order  & Number of photons per pulse\tabularnewline
\hline 
\hline 
11\textsuperscript{th}  & $\sim$1.4x10\textsuperscript{10}\tabularnewline
\hline 
13\textsuperscript{th}  & $\sim$8.4x10\textsuperscript{10}\tabularnewline
\hline 
14\textsuperscript{th}  & $\sim$8x10\textsuperscript{9}\tabularnewline
\hline 
16\textsuperscript{th}  & $\sim$2.3x10\textsuperscript{9}\tabularnewline
\hline 
17\textsuperscript{th}  & $\sim$3.2x10\textsuperscript{8}\tabularnewline
\hline 
\end{tabular}\caption{Number of photons per harmonic order per pulse at the source when
Ar is used as generating medium.}
\label{Table Number of Photons} 
\end{table}


\subsection{The Semiclassical analysis of HHG spectra and phases\label{subsec:The-Semiclassical-analysis-1}}

According to the semi-classical approach, the HHG from bi-circular
harmonic fields has different characteristics compared to the case
of linear monochromatic polarized driver fields. It was shown in \citep{Miloifmmodecheckselsevsfieviifmmodeacutecelsecfi2000}
that the main contribution to the harmonic emission comes from electrons
with Re(t)<$\nicefrac{T_{L}}{3}$ (where$T_{L}$is the fundamental
laser period). Here it should be stressed that because of the threefold
electric field which is raised by the superposition of the bi-circular
electric fields, only one trajectory mostly contribute during the
process of HHG, in contrast to the case of linear monochromatic drivers
where two trajectories contribute to harmonic emission, namely \textit{short}
and \textit{long} \citep{Kruse2010,Balcou1997,Bellini1998,Peatross1995}.

\begin{figure}[h]
\begin{centering}
\includegraphics[width=1\textwidth]{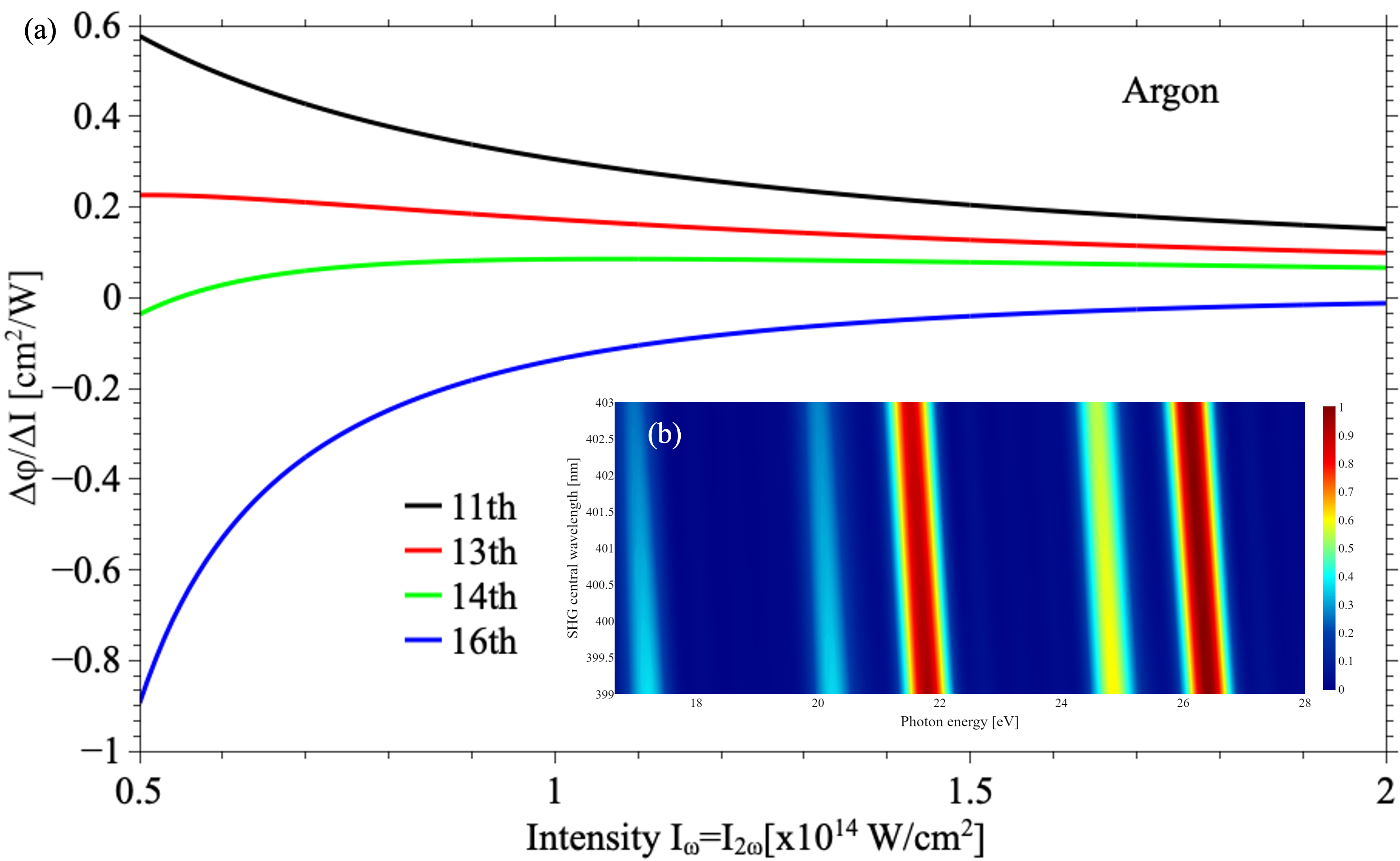} 
\par\end{centering}
\centering{}\caption{(a) Harmonic phase of the 3n$\pm$1 harmonics (n=4,5) as a function
of the intensity of the two components of the bi-circular counter
rotating fields. Parameters used in the calculations: laser central
wavelength 800 nm, 400nm second harmonic, pulse peak intensity $I_{\omega}=I_{2\omega}$,
Argon. (b) Harmonic spectra as as a function of the SHG central wavalength.
Parameters used in the calculations: fundamental central wavelength
$\lambda_{1}$ = 800 nm and tuning the second harmonic central wavelength
between 399 nm$\protect\leq\lambda_{2}\protect\leq$403 nm , pulse
peak intensity $I_{\omega}=I_{2\omega}$=1\texttimes 10$^{14}$ W/cm$^{2}$,
Argon.\label{fig:Harmonic-phase-of}}
\end{figure}

Fig. \ref{fig:Harmonic-phase-of} (a)represents the harmonic phase
of the 3n$\pm$1 harmonics (n = 4,5,6) as a function of the intensity
in the case where $\text{I}_{\omega}$=$\text{I}_{2\omega}$ of the
two components of the bi-circular counter rotating fields. Ar atomic
target is used as the generation medium. The result is that the slope
of the phase as a function of the intensity for each harmonic is $\emph{small}$.
This is indicative of a $\emph{strong collective response}$ when
loose focusing geometry is applied\citep{Fleischer2014,Kfir2015,Orfanos2019,Orfanos2020,Tzallas2005}.
For equal intensities of the laser field components $I_{\omega}=I_{2\omega}$
the cutoff law is $E_{max}$=1.2$I_{p}$+3.17$U_{p}$, where $U_{p}$=$U_{p\omega}$+$U_{p2\omega}$
\citep{Miloifmmodecheckselsevsfieviifmmodeacutecelsecfi2000,Hasovic2016}.

Using this semi-classical three step model \citep{Miloifmmodecheckselsevsfieviifmmodeacutecelsecfi2000,Lewenstein1994}
the HHG spectrum was also calculated for different SHG central wavelengths
(399 nm $\leq\lambda_{2}\leq$403 nm). It is shown here theoretically
(Fig. \ref{fig:Harmonic-phase-of}(b)) and verified experimentally
in Sec. \ref{subsec:Highly-elliptical-XUV-radiation} that the harmonic
photon energy follows a linear dependence on the SHG central wavelength.

\subsection{Probing the dynamical origin of the highly ellipticity harmonics
with TDDFT\label{subsec:Results-and-discussion}}

\begin{figure}[h]
\centering{} \includegraphics{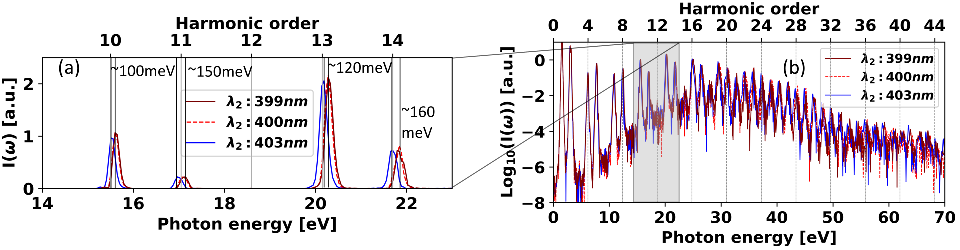}\caption{High-order harmonic spectra of an Argon atom in linear scale (a) and
logarithmic scale (b) obtained under the influence of BCCR laser field.
A wavelength of $\lambda_{1}$ = 800 nm for the fundamental laser
field is fixed and the wavelengths of the second harmonic field is
varied, i.e. $\lambda_{2}$ = 399, 400, and 403 nm. Energy shift in
the high harmonic spectra as a function of second harmonic laser field's
wavelength is shown in panel 'a'. \label{fig:HHG}}
\end{figure}

In this work, the peak intensity of the fundamental and the second
harmonic fields are taken as $1\times10^{14}W/cm^{2}$. As shown in
Fig. \ref{fig:Experimental-apparatus-for}(b) the total electric field
has a trefoil pattern. The harmonic generation and tunability of the
photon energy of the emitted harmonics in Argon atom is obtained by
combining a circular polarized light with a fundamental frequency
$\omega_{1}$ ($\lambda_{1}$ = 800 nm) with its counter-rotating
second harmonic $\omega_{2}$ ($\lambda_{2}$ = 399 nm or 400 nm or
401 nm or 403 nm). The harmonic radiation is plotted as the sum of
absolute square of the two polarization components (x and y), as shown
in linear and logarithmic-scale in Fig. \ref{fig:HHG}(a) and Fig.
\ref{fig:HHG}(b), respectively. The HHG spectrum in Fig. \ref{fig:HHG}(b)
displays a distinct structural peak in the lower frequency range of
15-30 eV.

We find that the emitted harmonic spectral features such as the HHG
intensity, polarization states of the harmonics, show strong dependence
on the wavelength and intensity ratio of the two driving field components
in the BCCR field. The dependence of the generated harmonic spectrum
on the SHG central wavelength, as obtained from our TDDFT analysis
show very good agreement with the theoretical calculations presented
in \ref{subsec:The-Semiclassical-analysis-1}. A maximum HHG central
energy shift of $\sim$150 meV is also observed by tuning the second
harmonic wavelength, as shown in Fig. \ref{fig:HHG}a supporting the
experimental results of \ref{subsec:Highly-elliptical-XUV-radiation}.



\begin{figure}[H]
\centering{}\includegraphics{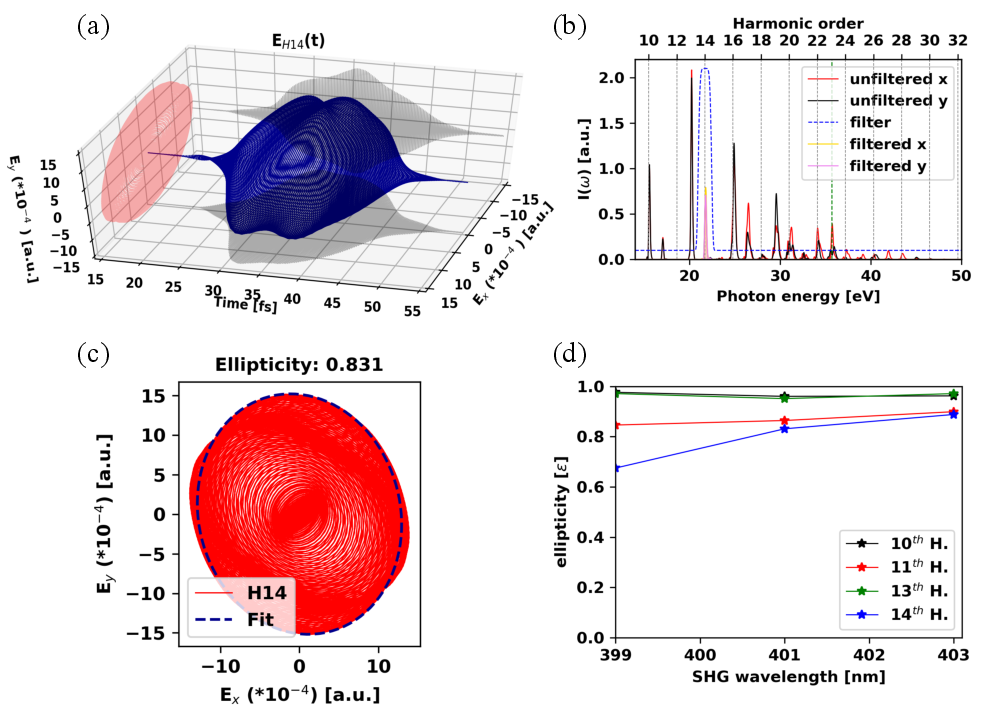}
\caption{High harmonic generation from Argon using a BCCR laser field with
$\lambda_{1}$ = 800 nm and $\lambda_{2}$ = 401 nm. (a) Temporal
evolution of the ellipticity of the emitted Harmonics super-Gaussian
filtered around 14$^{th}$ harmonic. (b) Computed HHG spectrum from
Argon, Super-Gaussian filtered (blue curve) around 14$^{th}$ harmonic.
The green colored vertical dashed line is the harmonic cutoff. (c)
The resulting harmonic field of the filtered harmonics (in red) is
sliced at its maximum and fitted to an ellipse (in blue).(d) Effect
of second harmonic frequency on the ellipticity of the harmonics\label{fig:ellip}}
\end{figure}

We investigate how the ellipticity of the emitted harmonics varies
as a function of second harmonic wavelength. For illustration purpose
we show for $\lambda_{2}~=~401$ nm the ellipticity of the 14th emitted
harmonics in Fig. \ref{fig:ellip}(a). As shown in Fig. \ref{fig:ellip}(b),a
super-Gaussian filter is applied around the 14th harmonic field (red
curve) of which is shown in Fig. \ref{fig:ellip}(c). A slice at the
peak the harmonic field is selected and an ellipse is fitted to it
(blue dashed curve). By taking the ratio of semi-minor and semi-major
axis, the ellipticity is calculated. The dependence of second harmonic
field wavelength on the ellipticity of the emitted harmonics is shown
in Fig. \ref{fig:ellip}(d).

We observe that by using the peak intensities of the fundamental field
($1\times10^{14}~W/cm^{2}$), the harmonic ellipticity corresponding
to the BCCR field induced HHG emission is significantly reduced for
all harmonic orders detectable in TOF spectrometer, indicating prominent
depolarization effects. In particular, by examining harmonic orders
10$^{th}$ to 14$^{th}$ (see Fig. \ref{fig:ellip}d), we find that
the calculated ellipticity varies from 0.65--0.98, same as in our
experimental results, without showing any predictable particular order,
while 10$^{th}$ and 13$^{th}$ harmonic show close-to-circularity
elliptic nature. Estimated degree of polarization of individual harmonic
orders exhibit similar gross features and magnitudes to that obtained
from our experimental results.

The difference of the harmonics ellipticity values derived from the
experimental data in Fig. \ref{fig:ellipticities} and those extracted
from the TDDFT calculations in Fig. \ref{fig:ellip}(d) can be attributed
to different factors related to the HHG process itself and also in
deviations from the optimal experimental conditions. Lou Barreau et
al. \citep{Barreau2018} showed that the breaking of the dynamical
symmetry of the three atto-pulses in the $\omega$ cycle (which is
the case of when the two color counter-rotating driving fields are
applied) would result in a decreased harmonic ellipticity, emission
of the 3q harmonic orders and possibly depolarization. This kind of
breaking can be induced either by the driving fields imperfect circular
polarizations\citep{Miloifmmodecheckselsevsfieviifmmodeacutecelsecfi2000,Fan2015}
, due to imperfect overlap\citep{JimenezGalan2017} or by the anisotropic
generating medium.\citep{Baykusheva2016,Yuan2018} Note that macroscopic
effects can further affect the polarization of the harmonics \citep{Antoine1996,Kfir2015}.

\section{Conclusions}

In conclusion, we report a method to produce tunable energetic highly-elliptical
XUV radiation. The approach is based on gas-phase HHG driven by two-color
bi-circular polarized fields, produced by a MAZEL--TOV-like device.
The approach is applied in the linearly polarized MW XUV beamline
at FORTH-IESL, under loose focusing conditions. By properly tuning
the central frequency of the second harmonic of the fundamental frequency,
the central frequency of the XUV HHG can be continuously tuned. By
performing polarimetric measurements ellipticities up to $\sim85$\%
at $\sim$22 eV were achieved. No clear dependence on the BBO angle
of rotation was observed. The energy per driving laser pulse, in the
spectral region $\HF$20 eV, was found to be in the range of E$^{XUV}$$\HF$400
nJ per laser pulse at the source. The maximum energy shift observed
was $\Delta$E$\HF$150 meV which is in good agreement with the theoretical
calculations.

Overall, by combining of state-of-the-art experiments, semi-classical
analysis and TDDFT simulations, we demonstrate efficient generation
and characterization of highly energetic and elliptic high harmonic
spectra from Argon, and also identify the dynamical origin and spectral,
temporal nature of the generated elliptic high order harmonics. We
find that the proposed technique to generate and tune the BCCR field
under suitable focusing conditions can result in highly energetic
(~400 nJ at the source) elliptically polarized harmonic spectra,
with a linear dependence of the central HHG energy on the central
wavelength of the SHG. While certain harmonics regions demonstrate
a high degree of ellipticity (almost tending to circularity, for example,
10$^{th}$ or 13$^{th}$ harmonics), other detectable high harmonics
demonstrate lower ellipticity. Similar features are also confirmed
with our TDDFT simulations. By suitably tuning and improving the focusing
conditions, we could achieve a high value of the energy content per
pulse of the highly-elliptical XUV emission (~400 nJ at the source),
which is much higher than previously reported results. Hence, employing
our approach based on short pulse driving laser, and by varying the
SHG central wavelength, it is possible to achieve spatially varying
elliptically polarized high harmonics that can be utilized in imaging
and spectroscopic applications in the materials \citet{Chatziathanasiou2019},
chemical, and nano sciences \citet{Kahaly2008}, as well as to probe
the chirality sensitive processes.
\begin{acknowledgments}
We acknowledge support of this work by the LASERLAB-EUROPE (EU’s Horizon
2020 Grant No. 871124), the IMPULSE project Grant No. 871161), the
Hellenic Foundation for Research and Innovation (HFRI) and the General
Secretariat for Research and Technology (GSRT) under grant agreements
{[}GAICPEU (Grant No 645){]} and NEA-APS HFRI-FM17-2668.ELI-ALPS is
supported by the European Union and co-financed by the European Regional
Development Fund (GINOP-2.3.6-15-2015-00001). The ELI-ALPS project
(GINOP-2.3.6-15-2015-00001) is supported by the European Union and
it is co-financed by the European Regional Development Fund. This
research has been supported by the IMPULSE project which receives
funding from the European Union Framework Programme for Research and
Innovation Horizon 2020 under grant agreement No 871161. SK and MUK
also acknowledges project No. 2019-2.1.13-TÉT-IN-2020-00059, which
has been implemented with support provided by the National Research,
Development and Innovation Fund of Hungary, and financed under the
2019-2.1.13-TÉT-IN funding scheme.SM and MUK would like to acknowledge
ELI-ALPS HPC administration for their support in providing computational
resources.

 \bibliographystyle{unsrt}
\phantomsection\addcontentsline{toc}{section}{\refname}\bibliography{EVassakis_et_al}
 \pagebreak{} 
\end{acknowledgments}

\section*{Supplementary Material}

\subsection{BBO Crystal spectral characteristics\label{subsec:BBO-Crystal-spectral}}

The BBO crystal is mounted on a high precision motorized rotational
stage and the spectral properties of the SHG can be tuned as a function
of the turning angle $\Delta$$\vartheta$ with respect to the propagation
axis of the 25-fs IR laser pulse. For a monochromatic fundamental
wave, the turning angle corresponds to the phase-mismatching angle
at the given fundamental wavelength. For a broadband laser pulse,
however, the turning angle means the variation in the phase-matching
angle, and the central SHG wavelength is subsequently tuned with respect
to the phase-matching angle. 
\begin{figure}[H]
\centering{}\includegraphics[width=1\textwidth]{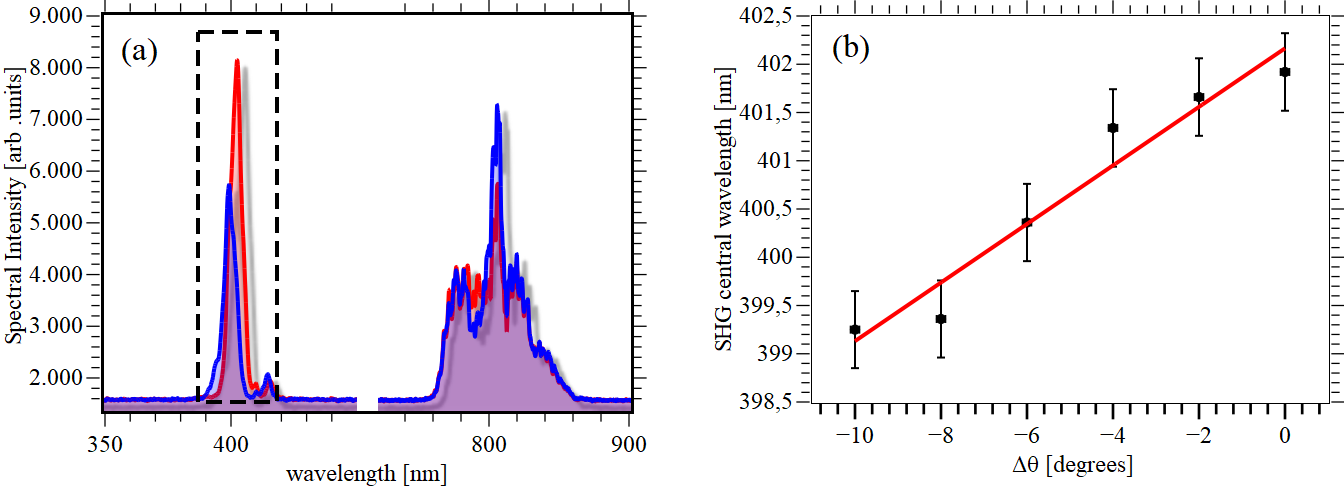}\caption{(a)Spectral distributions of the fundamental and SHG pulses for the
two extreme positions of the turning angle \ensuremath{\Delta}$\theta$
with the propagation axis of the 25-fs IR laser pulse.(b) SHG central
wavelength as a function of angle $\Delta$$\theta$. \label{fig:BBO-spectrum-vs}}
\end{figure}

It was verified experimentally (Fig. \ref{fig:BBO-spectrum-vs}(b))
that the central wavelength of the SHG pulse varies linearly with
the turning angle $\Delta$$\theta$, and the wavelength variation
$\Delta\lambda$$_{2}$ of the SHG pulse can be approximately expressed
as $\Delta$$\lambda$$_{2}$$\HF$0.3 $\Delta$$\theta$ ($-10^{o}\leq\Delta\theta\leq0^{o}$)
for the BBO (0.2 mm, cutting angle 29.20 for type I phase matching)
used in the experiment. The corresponding spectral distributions of
the fundamental and SHG for two extreme positions of the turning angle
\ensuremath{\Delta}$\theta$ are shown in Fig. \ref{fig:BBO-spectrum-vs}(a).

\subsection{Polarization - Reflective polarizer - Proof of equations used\label{subsec:Polarization---Reflective}}

The ellipticity of the highly elliptical XUV radiation produced through
HHG was extractedusing the XUV polarimeter. The analysis of the polarization
state of the incident light is achieved via a EUV reflective polarization
analyzer, which functions as a non-ideal linear polarizer/analyzer.
The polarization of light and its control are often described using
Jones vectors and matrices. Jones vectors describe the polarization
state of light. Jones matrices describe the effect of different optical
devices and processes on the polarization state. They are operators
that act on the incoming Jones vector and produce the out-coming Jones
vector. The electric field of a monochromatic (of angular frequency
$\omega$) plane electromagnetic wave propagating along the z axis
can be written as:

\begin{equation}
\left(\begin{array}{c}
E_{x}(t)\\
E_{y}(t)\\
0
\end{array}\right)=\left(\begin{array}{c}
E_{0x}(t)e^{i(kz-\omega t+\varphi_{x})}\\
E_{0y}(t)e^{i(kz-\omega t+\varphi_{y})}\\
0
\end{array}\right)=\left(\begin{array}{c}
E_{0x}(t)e^{i\varphi_{x}}\\
E_{0y}(t)e^{i\varphi_{y}}\\
0
\end{array}\right)e^{i(kz-\omega t)}\label{eq:12}
\end{equation}

where $k=\omega/c$.

The Jones vector is defined as the vector $\left(\begin{array}{c}
E_{0x}(t)e^{i\varphi_{x}}\\
E_{0y}(t)e^{i\varphi_{y}}\\
0
\end{array}\right)$, or simply

\begin{equation}
\left(\begin{array}{c}
E_{0x}(t)e^{i\varphi_{x}}\\
E_{0y}(t)e^{i\varphi_{y}}
\end{array}\right)\label{eq:13}
\end{equation}

\emph{Linearly polarized light} occurs when the direction of the electric
field remains constant, which means that the plane of polarization
(defined by the electric field and the direction of propagation) is
constant (or more precisely, parallel to the same plane). Linearly
polarized light at an angle $\theta$ with the x axis is described
by the Jones vector $\left(\begin{array}{c}
cos\theta\\
sin\theta
\end{array}\right)$ as $\varphi_{x}-\varphi_{y}$ must be zero or an integer multiple
of $\pi$.

\emph{Elliptically polarized light} (with circular polarization being
a special case) is generally described by the Jones vector

\begin{equation}
\left(\begin{array}{c}
E_{0x}(t)\\
E_{0y}(t)e^{i\epsilon}
\end{array}\right)\label{eq:14}
\end{equation}

The field vector describes an ellipse on the transverse plane (i.e.
the plane perpendicular to the direction of propagation). Special
case of elliptically polarized light is circularly polarized (CP)
light. Right hand CP light is described by the Jones vector $\frac{1}{\sqrt{2}}\left(\begin{array}{c}
1\\
-i
\end{array}\right)$ and left hand CP light is described by the Jones vector $\frac{1}{\sqrt{2}}\left(\begin{array}{c}
1\\
i
\end{array}\right)$.

Changes in the state of polarization can be achieved via several physical
processes, such as reflection, scattering, refraction, and transmission.
As previously mentioned, an optical element that transforms the polarization
state into another is described by the Jones matrix acting on the
incoming Jones vector: 
\begin{equation}
\left[\begin{array}{c}
E_{x}^{'}\\
E_{y}^{'}
\end{array}\right]=\left[\begin{array}{cc}
M_{11} & M_{12}\\
M_{21} & M_{22}
\end{array}\right]\left[\begin{array}{c}
E_{x}\\
E_{y}
\end{array}\right]\label{eq:15}
\end{equation}

\subsubsection{Jones matrix for a polarizer}

A linear polarizer does not affect the vibration direction of either
$E_{x}$ or $E_{y}$ . This means that in \ref{eq:15} $M_{12}=M_{21}=$0
and $\begin{array}{c}
E_{x}^{'}=p_{x}E_{x}\\
E_{y}^{'}=p_{y}E_{y}
\end{array}$, $0\leq p_{x},p_{y}\leq1$. Thus, the Jones matrix is written as:

\begin{equation}
J_{p}=\left[\begin{array}{cc}
p_{x} & 0\\
0 & p_{y}
\end{array}\right]\label{eq:16}
\end{equation}

where we set$M_{12}=p_{x}$ and $M_{21}=p_{y}$.

For an ideal horizontal polarizer, $J_{p}=\left[\begin{array}{cc}
1 & 0\\
0 & 0
\end{array}\right]$, while for an ideal vertical polarizer $J_{p}=\left[\begin{array}{cc}
0 & 0\\
0 & 1
\end{array}\right]$. For a non-ideal polarizer both $p_{x}$ and $p_{y}$ have non zero
values, but one is usually much smaller than the other. When a linear
polarizer is rotated by an angle $\frac{\theta}{}$ then the Jones
matrix needs to be accordingly rotated, using the rotation matrix.
R=$\left[\begin{array}{cc}
\cos\theta & \sin\theta\\
-\sin\theta & \cos\theta
\end{array}\right]$ , i.e.

\begin{equation}
J_{p}(\theta)=R^{-1}J_{p}R=\left[\begin{array}{cc}
\cos\theta & -\sin\theta\\
\sin\theta & \cos\theta
\end{array}\right]\left[\begin{array}{cc}
p_{x} & 0\\
0 & p_{y}
\end{array}\right]\left[\begin{array}{cc}
\cos\theta & \sin\theta\\
-\sin\theta & \cos\theta
\end{array}\right]=\left[\begin{array}{cc}
p_{x}\cos^{2}\theta+p_{y}sin^{2}\theta & (p_{x}-p_{y})\sin\theta\cos\theta\\
-(p_{x}-p_{y})\sin\theta\cos\theta & p_{x}\sin^{2}\theta+p_{y}\cos^{2}\theta
\end{array}\right]\label{eq:17}
\end{equation}

For the ideal case, for a horizontal polarizer, $p_{x}=1$ and $p_{y}=0$
and hence the Jones matrix becomes 
\begin{equation}
J_{p}(\theta)=\left[\begin{array}{cc}
\cos^{2}\theta & -\sin\theta\cos\theta\\
-\sin\theta\cos\theta & \sin^{2}\theta
\end{array}\right]\label{eq:18}
\end{equation}

Polarization by reflection can be described by the Jones matrix of
a non-ideal lineal polarizer.

\subsubsection{Malus’ law for an ideal and a real linear polarizer}

Let us assume that the incident electric field is linearly polarized
along the x axis, and can be described by the Jones vector $\left(\begin{array}{c}
1\\
0
\end{array}\right)$. Then, according to \ref{eq:18} the resulting Jones vector will
be given by $\left[\begin{array}{cc}
\cos^{2}\theta & -\sin\theta\cos\theta\\
-\sin\theta\cos\theta & \sin^{2}\theta
\end{array}\right]\left(\begin{array}{c}
1\\
0
\end{array}\right)=\left(\begin{array}{c}
\cos^{2}\theta\\
-\sin\theta\cos\theta
\end{array}\right)$.Where, amplitude of the emerging electric field will be therefore
equal to $E_{ox}\cos\theta\sqrt{\cos^{2}\theta+\sin^{2}\theta}$ and
hence the intensity of light ($\propto E^{2}$) will be given by 
\begin{equation}
I=I_{0}\cos^{2}\theta\label{eq:19}
\end{equation}

which is \emph{Malus law} for an ideal linear polarizer.

For a non-ideal polarizer, Malus law can be generalized using \ref{eq:17}
instead of \ref{eq:18}, as follows:

\begin{equation}
\left[\begin{array}{c}
E_{x}^{'}\\
E_{y}^{'}
\end{array}\right]=\left[\begin{array}{cc}
p_{x}\cos^{2}\theta+p_{y}sin^{2}\theta & (p_{x}-p_{y})\sin\theta\cos\theta\\
-(p_{x}-p_{y})\sin\theta\cos\theta & p_{x}\sin^{2}\theta+p_{y}\cos^{2}\theta
\end{array}\right]\left(\begin{array}{c}
1\\
0
\end{array}\right)=E_{0x}\left(\begin{array}{c}
p_{x}\cos^{2}\theta+p_{y}sin^{2}\theta\\
-(p_{x}-p_{y})\sin\theta\cos\theta
\end{array}\right)\label{eq:20}
\end{equation}

and 
\begin{equation}
I=I_{0}\left(p_{x}\cos^{2}\theta+p_{y}sin^{2}\theta\right)^{2}+\left((p_{x}-p_{y})\sin\theta\cos\theta\right)^{2}=I_{0}\left[p_{x}\cos^{2}\theta+p_{y}\sin^{2}\theta\right]\label{eq:21}
\end{equation}

This can also be re-written using the extinction ratio 
\begin{equation}
R=\frac{p_{x}^{2}}{p_{y}^{2}}\label{eq:22}
\end{equation}

as $I=I_{0}p_{x}\left[\cos^{2}\theta+R\sin^{2}\theta\right]$, which
becomes identical to \ref{eq:19} for $p_{x}=1$ and $p_{y}=0$.

\subsubsection{Malus law for a EUV reflective analyzer}

\ref{eq:21} can be further generalized, if the polarizer (as is the
case with the reflective analyzer developed in FO.R.T.H.-I.E.S.L.
and utilized for the polarimetric investigation presented in this
manuscript) also causes dephasing between the two components of the
electric field. In this case the Jones matrix can be written in the
more general form

\begin{equation}
J_{p}=\left[\begin{array}{cc}
p_{x}e^{i\psi_{x}} & 0\\
0 & p_{y}e^{i\psi_{y}}
\end{array}\right]\label{eq:23}
\end{equation}

We shall assume that the incident light is elliptically polarized
as in \ref{eq:14} which can be equivalently written in the form

\begin{equation}
\left(\begin{array}{c}
E_{0x}(t)e^{i\epsilon/2}\\
E_{0y}(t)e^{-i\epsilon/2}
\end{array}\right)\label{eq:24}
\end{equation}

Following the same methodology, we can write for the emerging field:

\begin{equation}
\left[\begin{array}{c}
E_{x}^{'}\\
E_{y}^{'}
\end{array}\right]=\left[\begin{array}{cc}
p_{x}e^{i\psi_{x}}\cos^{2}\theta+p_{y}e^{i\psi_{y}}sin^{2}\theta & (p_{x}e^{i\psi_{x}}-p_{y})\sin\theta\cos\theta\\
-(p_{x}e^{i\psi_{x}}-p_{y}e^{i\psi_{y}})\sin\theta\cos\theta & p_{x}e^{i\psi_{x}}\sin^{2}\theta+p_{y}e^{i\psi_{y}}\cos^{2}\theta
\end{array}\right]\left(\begin{array}{c}
E_{0x}(t)e^{i\epsilon/2}\\
E_{0y}(t)e^{-i\epsilon/2}
\end{array}\right)\label{eq:25}
\end{equation}

The intensity is then proportional to

\begin{equation}
I(\theta,\varepsilon)\propto E_{x}^{'2}+E_{y}^{'2}=p_{x}^{2}\left[E_{0x}^{2}(\cos^{2}\theta+R\sin^{2}\theta)+E_{0y}^{2}(R\cos^{2}\theta+\sin^{2}\theta)+E_{0x}^{2}E_{0y}^{2}(R-1)\sin2\theta\cos\varepsilon\right]\label{eq:26}
\end{equation}

where, R is given by \ref{eq:22} .

One can use \ref{eq:26} to derive the ratio $E_{0y}/E_{0x}$ for
the incident light, from the measured distribution of intensity as
a function of $\theta$, by noting the values of I at $\theta=0(I_{1})$
and $\theta=\pi/2(I_{2})$:

\begin{equation}
I_{1}\propto p_{x}^{2}\left[E_{0x}^{2}+RE_{0y}^{2}\right]\label{eq:27}
\end{equation}

\begin{equation}
I_{2}\propto p_{x}^{2}\left[E_{0x}^{2}R+E_{0y}^{2}\right]\label{eq:28}
\end{equation}

where $\beta=\frac{E_{0x}^{2}}{E_{0y}^{2}}$.

So

\begin{equation}
\frac{E_{0x}}{E_{0y}}=\sqrt{\frac{R-c}{cR-1}}\label{eq:29}
\end{equation}

,with $c\equiv contrast=\frac{I_{1}}{I_{2}}$ .

\subsection{TDDFT setup\label{TDDFT_suppl}}

The exponential of the Hamiltonian is calculated using LANCZOS method
given in Castro et al. \citep{Castro2004} and the XC potential is
represented by the local density approximation (LDA) \citep{Marques2012,Onida2002}.
All the calculations were performed using fully relativistic Hartwigsen,
Goedecker, and Hutter (HGH) pseudopotentials \citep{Hartwigsen1998}.
To understand the multielectron dynamics of Argon, single Ar atom
is placed in a parallelepiped simulation box of size 40 Bohr along
each of the three (x,y,z) cartesian directions, which includes 4 Bohr
of absorbing regions on either sides of the Argon. The absorbing regions
are treated using well-recognized mask function \citep{Koushki2018,Joachain2011,Mauger2019},
that ensures prevention of unphysical reflection of field-accelerated
electrons at the border of the simulation box. The real-space cell
was sampled along all the three directions by a grid spacing of 0.18
\text{Å}.

\subsection{The Semiclassical analysis of HHG spectra and phases\label{subsec:The-Semiclassical-analysis}}

In order to have an intuitive picture of the HHG by two color bi-circular
polarized fields, we performed calculations based on the semicalssical
approach where the theoretical framework and detailed analysis can
be found elsewhere \citep{Miloifmmodecheckselsevsfieviifmmodeacutecelsecfi2000}.
Here we present an abbreviated analysis which is based on Strong Field
Approximation (SFA)\citep{Lewenstein1994} adapted in the case of
these electric fields.

Within this model, the radiation at orders 3n$\pm1$ (with n=1,2,3,...),
emitted from a single atom exposed to a intense bi-chromatic driving
electric field \textbf{\textcolor{red}{E(t)}} with the associated
vector potential A(t) $=-\varint\mathbf{E}(t)dt$ can be fully characterized
by the Fourier transform of the time-dependent dipole moment.

\[
\mathbf{x}(\omega)=-i\int_{-\infty}^{\infty}\left\langle \varPsi(t)|\hat{\mathbf{x}}|\varPsi(t)\right\rangle e^{-i\omega t}dt=
\]

\[
=-i\int_{-\infty}^{\infty}dt{}_{r}\int_{-\infty}^{t_{r}}dt_{i}\int d^{3}\mathbf{p}\times[\mathbf{d}^{\ast}(\mathbf{p}-\mathbf{A}(t_{r}))\cdot\mathbf{E}(t_{i})]
\]

\begin{equation}
\mathbf{d}(\mathbf{p}-\mathbf{A}(t_{i}))\times e^{-iS(\mathbf{p},t_{i},t_{r})}+c.c.]\label{eq:ipole moment}
\end{equation}

In eq. \ref{eq:ipole moment}, $\left|\Psi(t)\right\rangle $ denotes
the time-dependent wave function of the electron that evolves according
to the Schrodinger's equation, and $\mathbf{\hat{x}=-r}$ is the dipole
operator. The temporal integration leads to an integration over the
three-dimensional momentum of the active electron $\mathbf{p}$, the
ionization time t$_{i}$, and recombination time t$_{r}$. Moreover,
$d(p)=\left\langle p\right|r\left|\psi\right\rangle $ is the dipole
matrix element, with the initial bound state $\left|\psi\right\rangle $
and a plane-wave state of momentum p, $\left|\psi\right\rangle =e^{-ipr}$
and for the case of the hydrogen-like atoms can be approximated as\citep{Lewenstein1994}:
\begin{equation}
\mathbf{d}(\mathbf{p})=i(\frac{1}{\pi\alpha})^{\nicefrac{3}{4}}\frac{\mathbf{p}}{\alpha}e^{-\frac{p^{2}}{2\alpha}}\label{eq:dipole moment}
\end{equation}
where $\alpha$ takes the value in the order of $I_{p}$ which is
the ionization potential of the atomic target.

In the exponent of eq. \ref{eq:ipole moment} the quantity S denotes
the quasi-classical action that the active electron experiences during
its excursion: 
\begin{equation}
S(\mathbf{p},t_{i},t_{r})=\int_{t_{i}}^{t_{r}}dt^{\prime\prime}(\frac{[p-A(t^{\prime\prime})]}{2}+I_{p})\label{eq:callsical action}
\end{equation}

Since the quasi-classical action in eq. \ref{eq:callsical action}
varies much faster than the other factors, it is not necessary to
solve the five dimensional integral for the dipole moment but we can
limit the evaluation of the integral over to the stationary point
of the classical action, $\nabla_{p}S(\mathbf{p_{s}},t_{i},t_{r})=0$,
where the quantity $\mathbf{p_{s}}$ is the stationary value of the
momentum, which can be obtained by equating the derivative of the
action in eq. \ref{eq:callsical action}with respect to $\mathbf{p}$
to zero. The complex phase is given by the equation:

\begin{equation}
\Theta(\mathbf{p_{s}},t_{i},t_{r})=\omega t_{r}-S(\mathbf{p_{s}},t_{i},t_{r})\label{eq:complex phase}
\end{equation}

Saddle-point approximation (SPA) requires the solution of the saddle-point
equations \citet{Nayak2019},obtained by equating the derivatives
of eq. \ref{eq:complex phase} with respect to $t_{i}$and $t_{r}$
to zero.

\begin{equation}
\nabla_{p}S(\mathbf{p_{s}},t_{si},t_{sr})=x(t_{sr})-x(t_{si})=0\label{eq:first saddle point equation}
\end{equation}

\begin{equation}
\frac{\partial\Theta(\mathbf{p_{s}},t_{i},t_{r})}{\partial t_{i}}\mid_{t_{si}}=\frac{1}{2}[\mathbf{p_{s}}-\mathbf{A}(t_{si})]^{2}+I_{p}\label{eq:second saddle point equation}
\end{equation}

\begin{equation}
\frac{\partial\Theta(\mathbf{p_{s}},t_{i},t_{r})}{\partial t_{i}}\mid_{t_{sr}}=-\frac{1}{2}[\mathbf{p_{s}}-\mathbf{A}(t_{si})]^{2}-I_{p}+\omega\label{eq:third saddle point equation}
\end{equation}
where all energies are expressed in terms of the photon energy, and
the right-hand side of eq. \ref{eq:third saddle point equation} denotes
the energy corresponding to photons with frequency $\omega$. Eq.
\ref{eq:first saddle point equation} indicates that the only relevant
electron trajectories are those where the electron leaves the atom
at time t$_{si}$ and returns at time t$_{sr}$. Equation \ref{eq:second saddle point equation}
describes energy conservation in the process of tunneling \citet{Nayak2019},
where the electron must have a negative kinetic energy at t$_{si}$,
leading to a complex value of t. Finally, eq. \ref{eq:third saddle point equation}
is the energy conservation law at the moment of recombination.

Using the stationary phase approximation \citet{Chatziathanasiou2019},
the Fourier transform of the dipole moment, x($\omega$) (eq. \ref{eq:ipole moment})
can be written as a coherent superposition of the contributions from
the complex electron quantum paths corresponding to the complex saddle-point
solutions $(\mathbf{p_{s}},t_{si},t_{sr})$. In the spirit of Feynman’s
path integrals \citep{Salieres2001} , can be expressed as:

\[
\mathbf{x}(\omega)=\underset{s}{\sum}\frac{i2\pi}{\sqrt{det(S^{\prime\prime})}}[\frac{\pi}{\varepsilon-i\frac{(t_{sr}-t_{si})}{2}}]^{\nicefrac{3}{2}}\mathbf{d^{\ast}}(\mathbf{p}_{s}-\mathbf{A}(t_{sr}))
\]

\begin{equation}
\times[\mathbf{\mathbf{E}(t_{si})\cdot d}(\mathbf{p_{s}}-\mathbf{A}(t_{si}))]\times e^{-iS(\mathbf{p},t_{si},t_{sr})+\omega t_{sr}}]
\end{equation}
where $det(S^{\prime\prime})$ is the determinant of the 2\texttimes 2
matrix of the second derivatives of $\Theta(\mathbf{p_{s}},t_{i},t_{r})$
(See eq. \ref{eq:complex phase}) with respect to t$_{i}$ and t$_{r}$
and the term $[\frac{\pi}{\varepsilon-i\frac{(t_{sr}-t_{si})}{2}}]^{\nicefrac{3}{2}}$
is related to the wave-packet spreading \citep{Sansone2004}. 
\end{document}